\documentclass[aps,pra,amsmath,amssymb,reprint,floatfix,superscriptaddress,showpacs]{revtex4-1}
\usepackage{graphicx}
\usepackage[bookmarks=false,colorlinks=true,linkcolor=blue,filecolor=blue,urlcolor=blue,citecolor=blue]{hyperref}
\usepackage{color}
\usepackage{soul}

\newcommand{\calA}{\mathcal{A}}
\newcommand{\calB}{\mathcal{B}}
\newcommand{\calC}{\mathcal{C}}

\newcommand{\ta}{\tilde{a}}
\newcommand{\tb}{\tilde{b}}
\newcommand{\tf}{\tilde{f}}
\newcommand{\tg}{\tilde{g}}
\newcommand{\tq}{\tilde{q}}
\newcommand{\tn}{\tilde{n}}
\newcommand{\tK}{\tilde{K}}
\newcommand{\tphi}{\tilde{\phi}}
\newcommand{\tchi}{\tilde{\chi}}

\newcommand{\brho}{\bar{\rho}}

\newcommand{\bfg}{\mathbf{g}}
\newcommand{\bfG}{\mathbf{G}}
\newcommand{\tbfg}{\tilde{\mathbf{g}}}

\newcommand{\bfz}{\mathbf{z}}
\newcommand{\bfK}{\mathbf{K}}
\newcommand{\bfJ}{\mathbf{J}}
\newcommand{\bfU}{\mathbf{U}}
\newcommand{\bfV}{\mathbf{V}}

\begin{document}

\title{Bridging coupled wires and lattice Hamiltonian for \\two-component bosonic quantum Hall states}

\author{Yohei Fuji}
\affiliation{Max-Planck-Institut f\"{u}r Physik komplexer Systeme, N\"{o}thnitzer Str. 38, 01187 Dresden, Germany}

\author{Yin-Chen He}
\affiliation{Max-Planck-Institut f\"{u}r Physik komplexer Systeme, N\"{o}thnitzer Str. 38, 01187 Dresden, Germany}

\author{Subhro Bhattacharjee}
\affiliation{International Centre for Theoretical Sciences, Tata Institute of Fundamental Research, Bangalore 560012, India}

\author{Frank Pollmann}
\affiliation{Max-Planck-Institut f\"{u}r Physik komplexer Systeme, N\"{o}thnitzer Str. 38, 01187 Dresden, Germany}

\date{\today}

\begin{abstract}
We investigate a model of hard-core bosons with correlated hopping on the honeycomb lattice in an external magnetic field by means of a coupled-wire approach. 
It has been numerically shown that this model exhibits at half filling the bosonic integer quantum Hall (BIQH) state, which is a symmetry-protected topological phase protected by the $U(1)$ particle conservation [Y.-C. He \textit{et al.}, Phys. Rev. Lett. \textbf{115}, 116803 (2015)]. 
By combining the bosonization approach and a coupled-wire construction, we analytically confirm this finding and show that it even holds in the strongly anisotropic (quasi-one-dimensional) limit. 
We discuss the stability of the BIQH phase against tunneling that break the separate particle conservations on different sublattices down to a global particle conservation. 
We further argue that a phase transition between two different BIQH phases is in a deconfined quantum critical point described by two copies of the $(2+1)$-dimensional $O(4)$ nonlinear sigma model with the topological $\theta$ term at $\theta=\pi$. 
Finally we predict a possible fractional quantum Hall state, the Halperin $(221)$ state, at $1/6$ filling. 
\end{abstract}

\maketitle

\section{Introduction}

In the last couple of decades, our understanding of gapped quantum phases of condensed matter has significantly advanced, especially with respect to topological phases that cannot be satisfactorily characterized by spontaneous symmetry breaking. 
These include two different classes of phases \cite{XChen10}: (i) topologically ordered phases, e.g., fractional quantum Hall phases, and (ii) symmetry-protected topological (SPT) phases, e.g., the Haldane phase in spin chains and topological band insulators.
The former class is characterized by long-range-quantum-entangled ground states that can support ``anyonic'' excitations with nontrivial braiding statistics (in two spatial dimensions) and associated topological degeneracy depending on the topology of the systems. 
Such phases remain well-defined as phases distinct from the trivial ones even in the absence of any symmetry.
SPT phases, on the other hand, have short-range-quantum-entangled ground states that remain well-defined only in the presence of certain symmetries.
For example, the Haldane phase is protected by the dihedral group of $\pi$ rotations about two orthogonal spin axes, time-reversal symmetry, or bond-centered inversion symmetry \cite{Pollmann10}.
While the SPT phases do not host any nontrivial bulk excitations, they typically possess nontrivial surface physics.
A variety of mathematical frameworks have been proposed to characterize and classify both the topologically ordered and SPT phases \cite{Levin05,Pollmann10, XChen10,YMLu12,Vishwanath13,XChen13a,Sule13,ZCGu14,LKong14,ChenjieWang15,ZBi15,Ringel15}. 
While free-fermion systems have by now been completely classified, the understanding of SPT phases in interacting fermionic and bosonic systems is an ongoing central topic of research.

A particularly interesting question in this regard is about how to stabilize exotic topological phases in systems of interacting bosons. 
Note that unlike fermions, which can host  topological phases even in noninteracting systems \cite{Haldane88,Kane05,Schnyder08}, noninteracting bosons usually condense and hence the interaction is essential to realize topological phases of bosons. 
This raises important questions regarding the nature of interactions in microscopic lattice models that can lead to such bosonic topological phases. 
However, the progress in controllably understanding these phases starting from lattice models is often hindered by the overarching problem of limitations in the present analytical approaches to deal with strongly interacting many-body systems, although the physics of strongly interacting bosonic models can in principle be very rich.
Under these circumstances, numerical approaches are of great help in understanding the fate of such systems. 
On the analytical side, a successful approach has been to construct nontrivial exactly solvable models, which then serve as ``parent Hamiltonians" to understand such correlated phases \cite{Kitaev03,Levin05,Hamma05,Kitaev06,Walker12,Levin11,Koch-Janusz13,XChen11,Levin12,Fidkowski13,Burnell14,Santos15,CHLin15}. 
An alternative way is the so-called projective construction \cite{Baskaran87,Baskaran88,Affleck88,Wen91,Sachdev92,FWang06,YMLu11,Grover13,YMLu14a,YMLu14b}. 
In this approach, the physical particles (microscopic degrees of freedom) of a model are broken up into a product of ``partons'', which are bosons or fermions. 
Then the ground state is obtained by projecting a mean-field ground state of those partons back onto the original Hilbert space. 

A complimentary approach, which can give a  more controlled way to access the physics of microscopic models than parton methods and is more general than parent Hamiltonian methods, is the coupled-wire construction \cite{Sondhi01,Kane02,Teo14,YMLu12,Mong14,Seroussi14,Vaezi14,Klinovaja14,Sagi14,Neupert14,Meng14,Meng15a,Gorohovsky15,Mross15a,Meng15b,Sagi15}. 
The general idea of this approach is as follows: We start from an array of wires (i.e., one-dimensional  systems), each of which can be described by using powerful techniques such as bosonization to yield an effective low-energy description.
The hoppings and interactions between wires are then incorporated to couple the low-energy modes of neighboring wires, yielding a description for the two-dimensional (2D) system. 
This prescription has been successfully implemented for integer and fractional quantum Hall states for which it was originally proposed \cite{Sondhi01,Kane02,Teo14}. 
In that context, at the single wire level one obtains Luttinger liquids in a magnetic field with left- and right-moving gapless modes. 
Inter-wire couplings mix these modes from different wires and open a gap successively from wire to wire, leading to a bulk gap and unpaired gapless modes---chiral edge states of the quantum Hall state---at the end wires. 
The above idea has recently been generalized to describe several other 2D topological phases \cite{YMLu12,Mong14,Seroussi14,Vaezi14,Klinovaja14,Sagi14,Neupert14,Meng14,Meng15a,Gorohovsky15} including the Kalmeyer-Laughlin chiral spin liquid on lattice spin systems \cite{Kalmeyer87,Gorohovsky15}, as well as several three-dimensional topological phases \cite{Mross15a,Meng15b,Sagi15}.
An obvious limitation of this approach is that quantitative prediction is restricted to highly anisotropic systems, that is weakly coupled wires. 
Nevertheless, this approach enables us to strictly keep track of the effects of interactions with the help of renormalization group and to study qualitative properties of the ground state beyond the perturbative regime unless some phase transition occurs by increasing the couplings between the wires. 
For example, the stability of a topologically ordered phase that supports the non-Abelian Fibonacci anyon, which is originally proposed by the coupled-wire construction in the quasi-one-dimensional limit \cite{Mong14}, is justified even in isotropic 2D lattice systems by numerical simulations \cite{Stoudenmire15}. 

In this paper, utilizing  bosonization techniques and the coupled-wire construction, we study two-component bosonic quantum Hall states realized on a lattice model.
We focus on a model of hard-core bosons interacting via correlated hoppings under background gauge fields on the honeycomb lattice, which was recently proposed in Ref.~\cite{YCHe15a} and shown to exhibit a bosonic integer quantum Hall (BIQH) state. 
The BIQH state is a bosonic SPT phase protected by a $U(1)$ symmetry related to the conservation of the total boson number \cite{XChen12a,YMLu12,XChen12b,Senthil13}. 
Although this state has been found in two-component Bose gases \cite{Furukawa13,YHWu13,Regnault13}, its lattice realizations have only been proposed and numerically studied quite recently \cite{YCHe15a,Sterdyniak15}. 
Here, we show that a spatially anisotropic limit of the model proposed in Ref.~\cite{YCHe15a} also exhibits the BIQH phase at half filling, which is possibly adiabatically connected to the one found by numerics \cite{YCHe15a}. 
Our calculations provide useful insights to the connection between the physics of the correlated hoppings in the model and the mutual flux attachment picture for the bosons as suggested in Ref.~\cite{Senthil13} for realizing the BIQH state;  due to this flux attachment, the condensation of the resulting composite bosons turns out to be not a superfluid but the BIQH phase with the Hall conductance $\sigma^{xy}=\pm 2$.
We then discuss the effect of tunnelings that break the $U(1)\times U(1)$ symmetry of charge conservation on two sublattices down to a simple $U(1)$ symmetry related to total charge conservation.
We also find indications that a possible continuous phase transition between different BIQH phases is described by two copies of the $(2+1)$-dimensional $O(4)$ nonlinear sigma model with the $\theta$ term at $\theta=\pi$. 
Further we extend our calculations to the case of $1/6$ filling for which we propose that the system stabilizes a bosonic fractional quantum Hall state---the so-called  Halperin (221) state \cite{Halperin83}.

The outline of this paper is as follows. 
In Sec.~\ref{sec:Model}, we introduce the model and its anisotropic deformation. 
We employ the standard bosonization procedure in Sec.~\ref{sec:Bosonization} and derive the low-energy effective theory of the model. 
In Sec.~\ref{sec:StableQHS}, utilizing the idea of coupled-wire construction, we find that the ground state corresponds to the BIQH phase for half filling and the Halperin (221) state for $1/6$ filling. 
The effect of inter-sublattice tunnelings and a possible effective description of the transition between the BIQH phases are also discussed. 
In Sec.~\ref{sec:Conclusion} we conclude with a short summary and discussion. 
Two appendices are devoted to various technical details. 

\section{Model} \label{sec:Model}

\begin{figure}
\includegraphics[clip,width=0.4\textwidth]{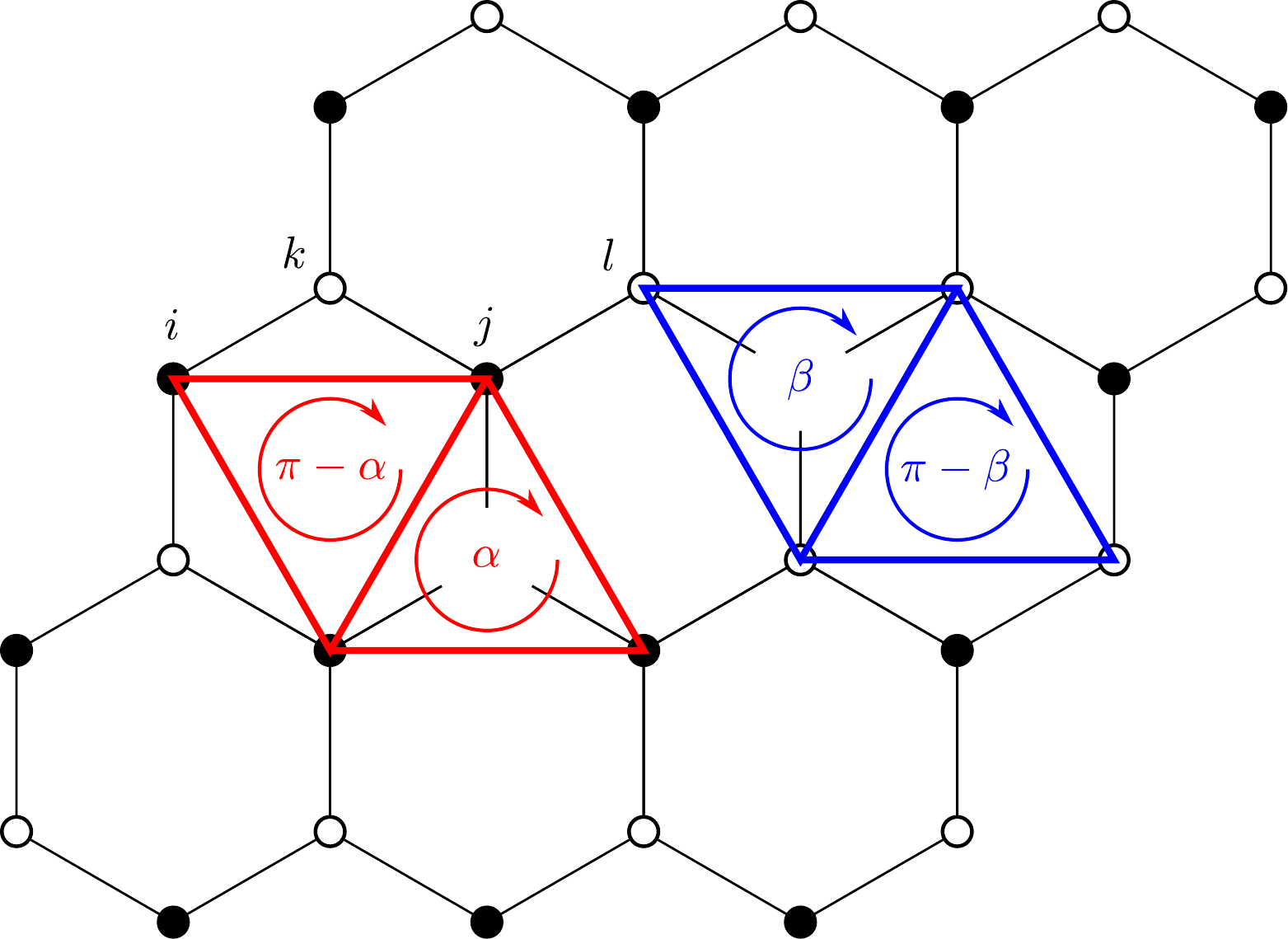}
\caption{(Color online) The model defined on the honeycomb lattice. 
The $A$ and $B$ sublattices are denoted by the filled and open circles, respectively. 
The red (blue) solid lines represent the correlated hoppings in the sublattice $A$ ($B$). 
On the $A$ sublattice, the upper triangles are pierced by the flux $\alpha$ while the lower ones by $\pi-\alpha$.  
On the $B$ sublattice, the lower triangles are pierced by the flux $\beta$ while the upper ones by $\pi-\beta$. 
}
\label{fig:Model}
\end{figure}

We start by describing the system of hard-core bosons on the honeycomb lattice (or a bilayer triangular lattice) in the presence of a background magnetic flux, which has been introduced in Ref.~\cite{YCHe15a}. 
The Hamiltonian is given by
\begin{align} \label{eq:OriginalHam}
H &= \sum_{\substack{\left< i, j; k \right> \\ (i,j \in A, k \in B)}} \left[ e^{i\calA_{ij}} a^\dagger_i a_j (2n^b_k-1) +\textrm{h.c.} \right] \nonumber \\
& +\sum_{\substack{\left< k,l; i \right>\\ (k,l \in B, i \in A)}} \left[ e^{i\calB_{kl}} b^\dagger_k b_l (2n^a_i-1) +\textrm{h.c.} \right],
\end{align}
where $a_i$ is the bosonic annihilation operator on site $i$ in the sublattice $A$, $b_i$ is that in the sublattice $B$, $n^a_i = a^\dagger_i a_i$, and $n^b_i = b^\dagger_i b_i$. 
These bosonic operators satisfy the hard-core constraint $(a_i)^2 = (b_i)^2 =0$. 
The correlated hoppings involve three sites denoted by $\left< i,j; k \right>$ as shown in Fig.~\ref{fig:Model}, where $i$ and $j$ are next-nearest-neighboring sites belonging to the same sublattice while $k$ represents the site between $i$ and $j$ that belongs to the other sublattice. 
The fluxes $\calA$ and $\calB$ are assigned as follows: 
\begin{align}
\begin{split}
\sum_{i,j \in \bigtriangleup (i,j \in A)} \calA_{ij} &= \alpha, \\
\sum_{i,j \in \bigtriangledown (i,j \in A)} \calA_{ij} &= \pi-\alpha, \\
\sum_{k,l \in \bigtriangleup (k,l \in B)} \calB_{kl} &= \pi-\beta, \\
\sum_{k,l \in \bigtriangledown (k,l \in B)} \calB_{kl} &= \beta, 
\end{split}
\end{align}
where all the pairs of site indices run in the clockwise around each triangle as shown in Fig.~\ref{fig:Model}. 
Also $\mathcal{A}_{ij}=-\mathcal{A}_{ji}$ and $\mathcal{B}_{kl}=-\mathcal{B}_{lk}$.
Thus, each hexagon is pierced by a uniform flux $\pi$ and each triangle on the sublattice is further threaded by a staggered flux: $\alpha$ for the up triangles while $\pi-\alpha$ for the down ones that are associated with the $A$ sublattice, and $\beta$ for the down triangles while $\pi-\beta$ for the up ones associated with the $B$ sublattice. 
The hopping between sites $i$ and $j$ on the $A$ sublattice changes its sign, depending on whether the intermediate site $k$ on the $B$ sublattice is occupied on not, and similarly for the hopping on the $B$ sublattice. 
This extra sign can be interpreted as an additional mutual flux seen by one species of bosons due to the other species. 
This resembles a mutual flux attachment proposed by Senthil and Levin \cite{Senthil13} to realize a BIQH phase; one species of boson binds one flux quantum of the other species by a mutual Chern-Simons term. 
As we will see below through the coupled-wire construction, our model indeed favors those ``mutual composite bosons'' as fundmental degrees of freedom rather than the standard bosons. 
These naturally raise the BIQH and Halperin $(221)$ states as candidates of ground state for given fillings.

Apart from the various lattice symmetries, in the above model, the two flavors of bosons associated with the two sublattices are independently conserved, i.e. there is a $U(1)_a \times U(1)_b$ symmetry where $a$ and $b$ are the two boson flavors introduced earlier. 
Alternatively we may state that the above symmetry can be cast as separate conservation of the total number of bosons (both $a$ and $b$) and the difference between the two sublattices. 
The conservation of the total boson charge (we use number and charge interchangeably) is represented by $U(1)_c$, while the conservation of their difference, which we denote as pseudospin, by $U(1)_s$. 
In terms of charge and pseudospin, the symmetry of the model is thus $U(1)_c \times U(1)_s$. 

This model [Eq.~\eqref{eq:OriginalHam}] was introduced in Ref.~\cite{YCHe15a} and studied numerically by three of the present authors. 
The hopping between nearest-neighboring sites, which is omitted from Eq.~\eqref{eq:OriginalHam}, destroys the conservation of the pseudospin $U(1)_s$ symmetry such that the remaining symmetry is only $U(1)_c$ (apart from the lattice symmetries). 
The numerical results confirmed that the BIQH phase is robust to such perturbations, as the SPT phase can be protected by the $U(1)_c$ symmetry alone.
We will introduce these terms and discuss their effect in Sec.~\ref{sec:SBTerm}. 

\subsection{Mapping to the array of chains}

\begin{figure}
\includegraphics[clip,width=0.45\textwidth]{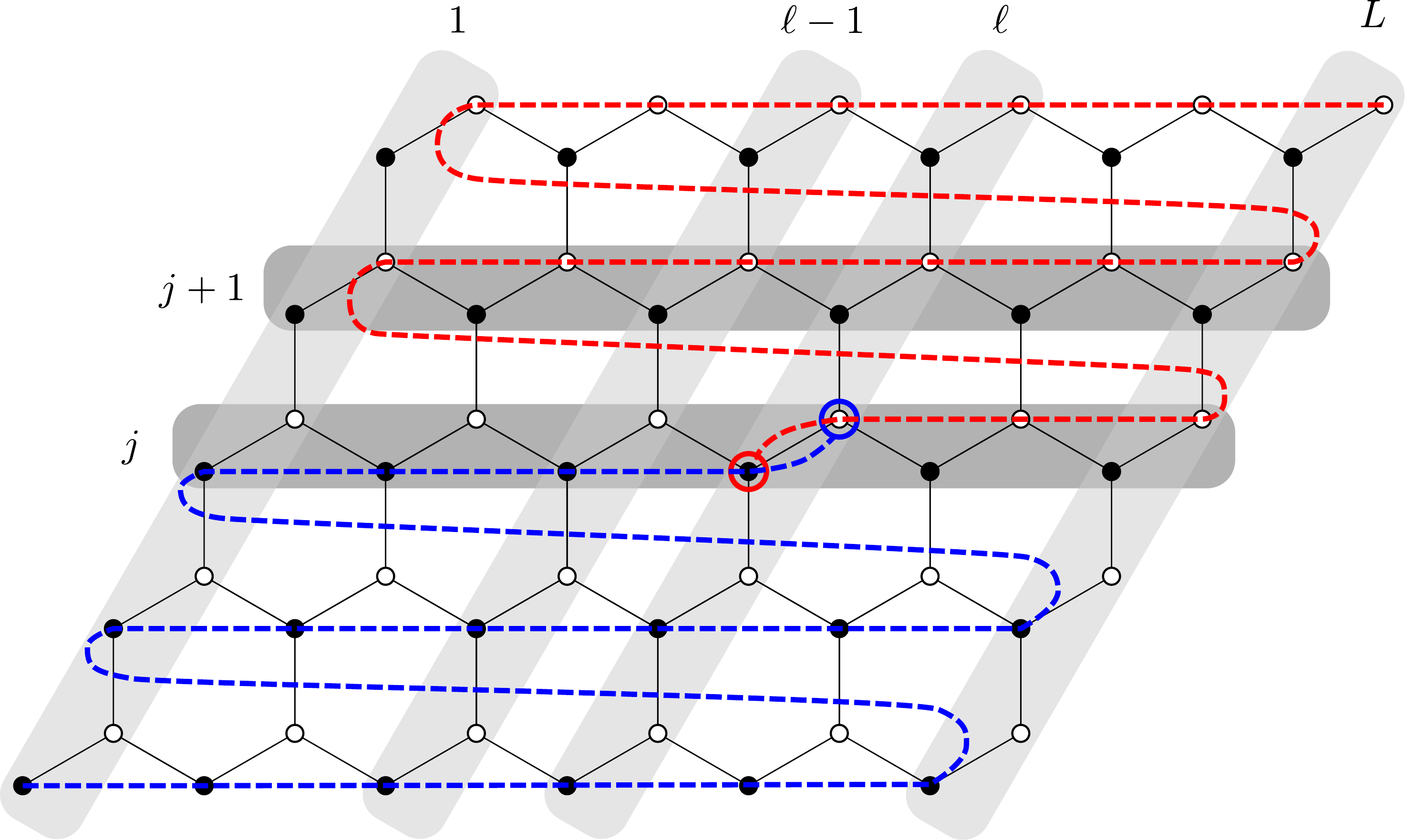}
\caption{(Color online) Redefinition of the site index for the honeycomb lattice to be viewed as coupled chains. 
The red (blue) dashed line represents the path of the string operator for $\ta_{j,\ell}$ ($\tb_{j,\ell}$).}
\label{fig:ChainHoneycomb}
\end{figure}

In order to treat the Hamiltonian in Eq.~\eqref{eq:OriginalHam} within the coupled-wire approach, we need to write it as a quasi-one-dimensional system. 
To this end, it is convenient to rewrite the Hamiltonian in the following form, 
\begin{align} \label{eq:Ham}
H =& \ H_0 +H_1, \\
\label{eq:ChainHamCorr}
H_0 =& \ t\sum_{j=1}^N \sum_{\ell=1}^L \Bigl[ e^{i\calA^{(j,\ell)}_{(j,\ell+1)}} a^\dagger_{j,\ell} a_{j,\ell+1} (2n^b_{j,\ell}-1) \nonumber \\
& +e^{i\calB^{(j,\ell)}_{(j,\ell+1)}} b^\dagger_{j,\ell} b_{j,\ell+1} (2n^a_{j,\ell+1}-1) + \textrm{h.c.} \Bigr], \\
H_1 =& \ t' \sum_{j=1}^N \sum_{\ell=1}^L \Bigl[ e^{i\calA^{(j,\ell)}_{(j+1,\ell-1)}} a^\dagger_{j,\ell} a_{j+1,\ell-1} (2n^b_{j,\ell-1}-1) \nonumber \\
& +e^{i\calA^{(j,\ell)}_{(j+1,\ell)}} a^\dagger_{j,\ell} a_{j+1,\ell} (2n^b_{j,\ell}-1)  \nonumber \\ 
& +e^{i\calB^{(j,\ell)}_{(j+1,\ell-1)}} b^\dagger_{j,\ell} b_{j+1,\ell-1} (2n^a_{j+1,\ell}-1) \nonumber \\
& +e^{i\calB^{(j,\ell)}_{(j+1,\ell)}} b^\dagger_{j,\ell} b_{j+1,\ell} (2n^a_{j+1,\ell}-1) +\textrm{h.c.} \Bigr],
\end{align}
where the ``chain'' index $j$ and ``site'' index $\ell$ are assgined as in Fig.~\ref{fig:ChainHoneycomb}. 
This mapping to the array of chains and the ``Jordan-Wigner-like'' transformation defined below are somewhat similar to those introduced to solve the Kitaev-honeycomb model \cite{Kitaev06,HDChen08}. 
Here we put the model on a cylinder with the length $N$ and the circumference $L$, such that the respective sites in each chain $\ell=1$ and $\ell=L+1$ are identified. 
Here, we have explicitly introduced the hopping amplitudes $t$ and $t'$.  
At $t=t'=1$, the model is spatially isotropic and we recover Eq.~\eqref{eq:OriginalHam}. 
Throughout this paper, we assume that $t'$ is perturbatively smaller than the chain hopping $t$. 

\subsection{Jordan-Wigner-like transformation}

Although the chain Hamiltonian \eqref{eq:ChainHamCorr} is not in the quadratic form, it can be transformed into a simple hopping Hamiltonian by the following Jordan-Wigner-like transformation. 
To implement the transformation, we introduce new operators $\tilde a_{j,\ell}$ and $\tilde b_{j,\ell}$ on the $A$ and $B$ sublattice respectively such that 
\begin{align} \label{eq:TransBoson}
\begin{split}
\ta_{j,\ell} &= K^b_{j,\ell} a_{j,\ell}, \\
\tb_{j,\ell} &= K^a_{j,\ell} b_{j,\ell}, 
\end{split}
\end{align}
where $K^s_{j,\ell}$, $s=a,b$ are the string operators defined by 
\begin{align} \label{eq:StringExp}
\begin{split}
K^a_{j,\ell} &= \exp \left( i\pi \sum_{j'<j} \sum_{\ell'=1}^L n^a_{j',\ell'} +i\pi \sum_{\ell' \leq \ell} n^a_{j,\ell'} \right), \\
K^b_{j,\ell} &= \exp \left( i\pi \sum_{j'>j} \sum_{\ell'=1}^L n^b_{j',\ell'} +i\pi \sum_{\ell' \geq \ell} n^b_{j,\ell'} \right). 
\end{split}
\end{align}
The paths of the string operators are depicted in Fig.~\ref{fig:ChainHoneycomb}.  
As shown in the figure, the string attached to the operator $a_{j,\ell}$ comes from the upper end of the cylinder, sweeps sites on the sublattice $B$ with winding the cylinder, and ends up at the site $(j,\ell)$. 
On the other hand, the string attached to $b_{j,\ell}$ comes from the lower end and sweeps all the sites on the sublattice $A$ before reaching the site $(j,\ell)$. 
Thus, the two strings wind the cylinder in opposite directions with respect to each other when we look down from either end of the cylinder. 
At this stage we introduce a more symmetric form of the string operators [Eq.~\eqref{eq:StringExp}] for later convenience (see also Ref.~\cite{Giamarchi}), as 
\begin{align} \label{eq:StringCos}
\begin{split}
K^a_{j,\ell} &= \cos \left( \pi \sum_{j'<j} \sum_{\ell'=1}^L n^a_{j',\ell'} +\pi \sum_{\ell' \leq \ell} n^a_{j,\ell'} \right), \\
K^b_{j,\ell} &= \cos \left( \pi \sum_{j'>j} \sum_{\ell'=1}^L n^b_{j',\ell'} +\pi \sum_{\ell' \geq \ell} n^b_{j,\ell'} \right). 
\end{split}
\end{align}
It is easy to check that the two formulas \eqref{eq:StringExp} and \eqref{eq:StringCos} are equivalent since $n^s_{j,\ell}$ only takes integer values $0$ or $1$. 

The new operators $\ta$ and $\tb$ still obey the bosonic statistics by the above choice of the strings.
If the sting operator $K^a_{j,\ell}$ involves $n^a_{j',\ell'}$, it produces a minus sign when $K^a_{j,\ell}$ goes over the operator $a_{j',\ell'}$. 
Similarly, the string operator $K^b_{j,\ell}$ produces a minus sign when it involves $n^b_{j',\ell'}$ and goes over $b_{j',\ell'}$. 
However, when $\ta_{j,\ell}$ goes over $\tb_{j',\ell'}$, this process always produces a minus sign zero or two times; thus they obey the bosonic statistics. 
At this point it is interesting to note that different choices of the string operators give the particle operators obeying mutual or genuine fermionic statistics; the latter is achieved by the standard 2D Jordan-Winer transformation \cite{Fradkin89}. 
However, we also note that no matter how we choose the Jordon-Wigner string, we end up with the same conclusion as long as one carefully keeps track of the particle statistics in the bosonization procedure. 

Using the transformation~\eqref{eq:TransBoson}, the chain Hamiltonian is simply written as 
\begin{align} \label{eq:ChainHam}
H_0 = -t \sum_{j=1}^N \sum_{\ell=1}^L \left[ e^{i\calA_\parallel} \ta^\dagger_{j,\ell} \ta_{j,\ell+1} +e^{i\calB_\parallel} \tb^\dagger_{j,\ell} \tb_{j,\ell+1} +\textrm{h.c.} \right]. 
\end{align}
where we have parametrized the gauge fields to be homogeneous along the chain: 
\begin{align}
\calA^{(j,\ell)}_{(j,\ell+1)} \equiv \calA_\parallel, \hspace{10pt} \calB^{(j,\ell)}_{(j,\ell+1)} \equiv \calB_\parallel. 
\label{eq:para}
\end{align}
This choice allows us to take a simple continuum limit along the chains as we perform below.
In Eq.~\eqref{eq:ChainHam}, the density operators in the correlated hoppings are canceled with the string operators due to the relation, 
\begin{align} \label{eq:DensityToString}
2n^s_{j,\ell}-1 = -e^{i\pi n^s_{j,\ell}}. 
\end{align}
Expressing the original bosonic operators in terms of $\ta$ and $\tb$, the interchain coupling Hamiltonian $H_1$ [in Eq.~\eqref{eq:Ham}] can be written as
\begin{align} \label{eq:IntHamString}
H_1 =& -t' \sum_{j=1}^N \sum_{\ell=1}^L \Bigl[ e^{i\calA^{(j,\ell)}_{(j+1,\ell-1)}} \ta^\dagger_{j,\ell} \ta_{j+1,\ell-1} \tK^b_{j,\ell-1} \tK^b_{j+1,\ell-1} \nonumber \\
& +e^{i\calA^{(j,\ell)}_{(j+1,\ell)}} \ta^\dagger_{j,\ell} \ta_{j+1,\ell} \tK^b_{j,\ell+1} \tK^b_{j+1,\ell} \nonumber \\
& +e^{i\calB^{(j,\ell)}_{(j+1,\ell-1)}} \tb^\dagger_{j,\ell} \tb_{j+1,\ell-1} \tK^a_{j,\ell} \tK^a_{j+1,\ell} \nonumber \\
& +e^{i\calB^{(j,\ell)}_{(j+1,\ell)}} \tb^\dagger_{j,\ell} \tb_{j+1,\ell} \tK^a_{j,\ell} \tK^a_{j+1,\ell-1} +\textrm{h.c.} \Bigr],
\end{align}
where 
\begin{align} \label{eq:BosonOpTrans}
\begin{split}
a_{j,\ell} &= \tK^b_{j,\ell} \ta_{j,\ell}, \\
b_{j,\ell} &= \tK^a_{j,\ell} \tb_{j,\ell}, 
\end{split}
\end{align}
and
\begin{align} \label{eq:StringTilde}
\begin{split}
\tK^a_{j,\ell} &= \cos \left( \pi \sum_{j'<j} \sum_{\ell'=1}^L \tn^a_{j',\ell'} +\pi \sum_{\ell' \leq \ell} \tn^a_{j,\ell'} \right), \\
\tK^b_{j,\ell} &= \cos \left( \pi \sum_{j'>j} \sum_{\ell'=1}^L \tn^b_{j',\ell'} +\pi \sum_{\ell' \geq \ell} \tn^b_{j,\ell'} \right). 
\end{split}
\end{align}
In the above expression, we have used $\tn^s_{j,\ell} = n^s_{j,\ell}$ and the density operators have been absorbed into the string operators by using Eq.~\eqref{eq:DensityToString}. 
A price to pay of the Jordan-Wigner-like transformation is that we need to treat the nonlocal interchain interactions involving the strings. 
However, those interactions can be converted to local interactions after the bosonization procedure, as we will see in the next section. 

We note that one can proceed to the bosonization analysis without incorporating the density operators into the strings. 
This will lead to a slightly different continuum expression for the interaction Hamiltonian. 
However the resulting physics is not changed whichever we choose, as shown in Appendix~\ref{app:AlternativeHam}. 
In the following, we consider the interaction Hamiltonian of the form \eqref{eq:IntHamString}. 

\section{Bosonization} \label{sec:Bosonization}

Following the standard bosonization techniques \cite{Giamarchi,Fradkin}, we here derive the effective low-energy theory for the Hamiltonian given by Eq.~\eqref{eq:Ham}.
We take the continuum limit with respect to the site index $\ell$, while the chain index $j$ is kept discrete. 
Thus we introduce a continuous variable $x=\ell a_0$ with $a_0$ begin the lattice spacing on the sublattices. 
Since the chain Hamiltonian \eqref{eq:ChainHam} is nothing but the hopping Hamiltonian of hard-core bosons, or equivalently the spin-1/2 XX chain, it can be described by an array of two-component Luttinger liquids, 
\begin{align} \label{eq:ChainBoson}
H_0 \sim \sum_{j=1}^N \sum_{s=a,b} \frac{v}{2\pi} \int dx \left[ (\partial_x \varphi^s_j)^2 +(\partial_x \theta^s_j)^2 \right],
\end{align}
where $v=2ta_0 \sin(\pi \brho a_0)$ and $\brho$ is the mean density of the boson. 
Here and hereafter, we assume that there is no modulation of boson density among chains, leading to the same mean density $\brho$ on both the $A$ and $B$ sublattices.
The bosonic fields $\theta^s_j(x)$ and $\varphi^s_j(x)$ obey the commutation relations, 
 \begin{align} \label{eq:BosonFieldComm}
[\theta^s_j(x), \varphi^{s'}_{j'}(x')] = i\pi \delta_{ss'} \delta_{jj'} \Theta(x-x'), 
\end{align}
where $\Theta(x)$ is a step function such that $\Theta(x)=1$ for $x>0$ while $\Theta(x)=0$ for $x<0$. 
The particle operators are bosonized as 
\begin{align} \label{eq:BosonMutualOp}
\begin{split}
\ta_{j,\ell} &\sim e^{-i\varphi^a_j(x) -i\calA_\parallel x/a_0} \left[ c_1 +c_2 \cos (2\pi \brho x +2\theta^a_j(x)) \right], \\
\tb_{j,\ell} &\sim e^{-i\varphi^b_j(x) -i\calB_\parallel x/a_0} \left[ c_1 +c_2 \cos (2\pi \brho x +2\theta^b_j(x)) \right],
\end{split}
\end{align}
where $c_{1,2}$ are nonuniversal constants. 
The density operators are bosonized as
\begin{align} \label{eq:DensityBoson}
\tn^s_{j,\ell} &\sim \brho a_0 +\frac{a_0}{\pi} \partial_x \theta^s_j(x) +\frac{1}{\pi} \cos(2\pi \brho x +2\theta^s_j(x)). 
\end{align}
We note that in Eqs.~\eqref{eq:BosonMutualOp} and \eqref{eq:DensityBoson}, we only kept the most relevant terms. 
In general, they contain less relevant terms, which are the vertex operators of $2q \theta^s_j(x)$ with integers $q>1$, but those operators do not enter the following analysis. 

We then consider the bosonized expressions of the string operators \eqref{eq:StringTilde}. 
For the arguments of the cosine terms of the string operators, by replacing the sum over $\ell$ by the integral over $x$, we may have
\begin{align}
&\sum_{j'<j} \sum_{\ell'=1}^{L} \tn^a_{j',\ell'} +\sum_{\ell' \leq \ell} \tn^a_{j,\ell'} \nonumber \\
&\sim \sum_{j'<j} \int_0^{La_0} dy \left( \brho +\frac{1}{\pi} \partial_y \theta^a_{j'} \right) +\int_0^x dy \left( \brho +\frac{1}{\pi} \partial_y \theta^a_j \right). 
\end{align}
for $\tK^a_{j,\ell}$, and
\begin{align}
&\sum_{j'>j} \sum_{\ell'=1}^L \tn^b_{j',\ell'} +\sum_{\ell' \geq \ell} \tn^b_{j,\ell'} \nonumber \\
&\sim \sum_{j'>j} \int_0^{La_0} dy \left( \brho +\frac{1}{\pi} \partial_y \theta^b_{j'} \right) +\int_x^{La_0} dy \left( \brho +\frac{1}{\pi} \partial_y \theta^b_j \right). 
\end{align}
for $\tK^b_{j,\ell}$. 
In the above expressions, terms proportional to $\brho$ yield an extra overall phase factor of $\brho L$. 
Such factors can be safely neglected by taking $L$ such that $\brho L a_0=2 n$ (where $n$ is an integer).
Thus we can write 
\begin{align} \label{eq:StringBoson}
\begin{split}
\tK^a_{j,\ell} &\sim \cos \left( \pi \brho x +\theta^a_j(x) +\pi \sum_{j' < j} N^a_{j'} \right), \\
\tK^b_{j,\ell} &\sim \cos \left( \pi \brho x +\theta^b_j(x) +\pi \sum_{j' \geq j} N^b_{j'} \right), 
\end{split}
\end{align}
where the coordinate-free operator $N^s_j$ is defined as \cite{Teo14}
\begin{align}
N^s_j = \frac{1}{\pi} \int_0^{La_0} dx \ \partial_x \theta^s_j(x), 
\end{align}
which is nothing but the spatially uniform twist in $\theta_j^s(x)$, that is the zero-mode part of the mode expansion of $\theta^s_j(x)$. 
$N^s_j$ obeys the commutation relation
\begin{align}
[N^s_j, \varphi^{s'}_{j'}(x)] = i\delta_{ss'} \delta_{jj'}.
\end{align}
Hence its eigenvalues are integers. 
Thus this operator produces a minus sign whenever $\tK^a_{j,\ell}$ goes over $\ta_{j',\ell}$ with $j' < j$ since 
\begin{align}
e^{i\pi N^a_{j,0}} e^{-i\varphi^a_j} &= e^{-i\varphi^a_j} e^{i\pi N^a_j} e^{\pi [N^a_j, \varphi^a_j]} \nonumber \\
&= -e^{-i\varphi^a_j} e^{i\pi N^a_j}. 
\end{align}
A similar anticommuting property also holds between $\tK^b_{j,\ell}$ and $\tb_{j',\ell}$ for $j' \geq j$. 
In combination with the commutation relation \eqref{eq:BosonFieldComm}, this relation ensures that the original bosonic operators $a$ and $b$ commute with each other. 

To be precise, for the bosonization formulas of the string operators \eqref{eq:StringBoson}, one should take a sum over any vertex operator $e^{in\theta^s_j}$ with odd $n$, as the lattice string operators \eqref{eq:StringExp} do not change under the replacement of $i\pi$ by $in\pi$ with odd $n$. 
However, as we will see in Sec.~\ref{sec:StableQHS}, interactions involving the vertex operators with large $n$ are generally irrelevant in low energy and at a commensurate filling not too far from half filling. 
In the following analysis, such higher harmonics do not have any crucial role. 
Therefore, we only keep $e^{in\theta^s_j}$ with the smallest odd integers $n=\pm 1$ as in Eq.~\eqref{eq:StringBoson}. 

\begin{figure*}
\includegraphics[clip,width=0.9\textwidth]{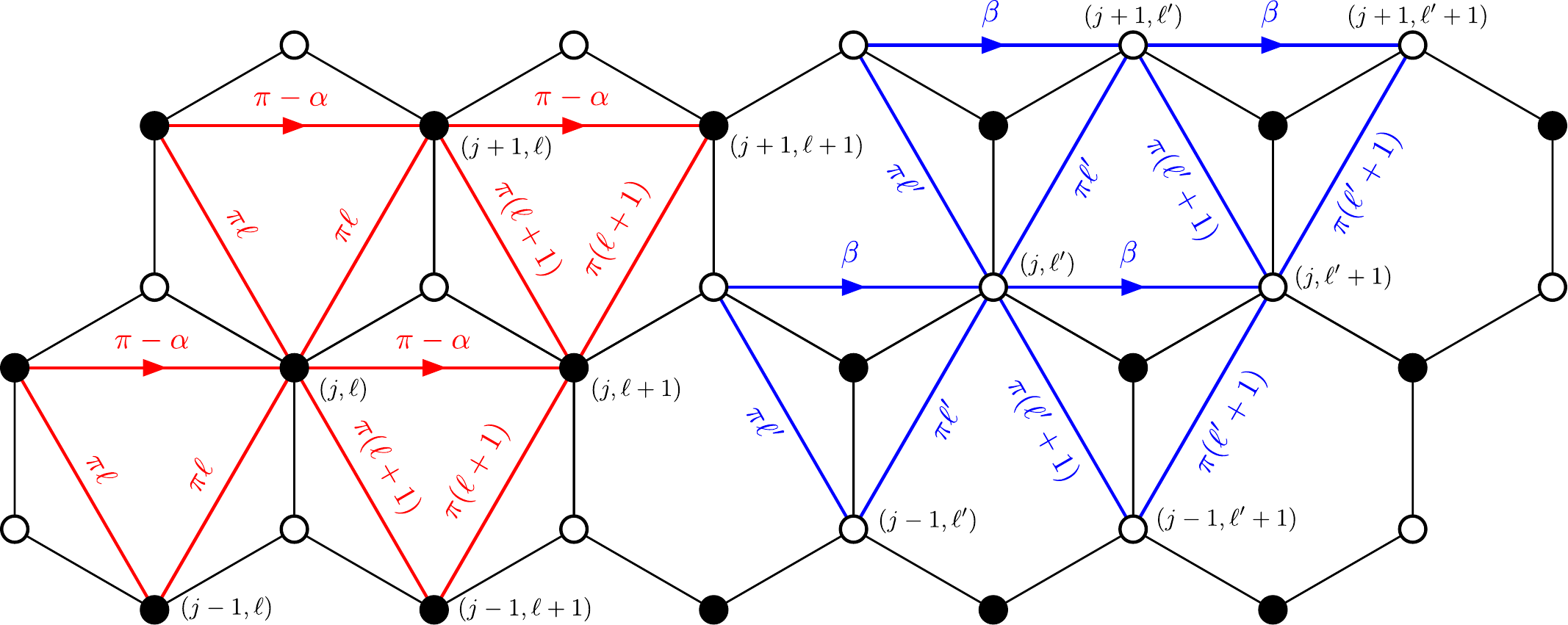}
\caption{(Color online) A choice of the phases for the correlated hopping terms, which gives Eqs.~\eqref{eq:FixedFlux} and \eqref{eq:FixedFlux2}. 
The red (blue) arrows or lines indicate the phases of the correlated hoppings on the $A$ ($B$) sublattice.}
\label{fig:FixedFlux}
\end{figure*}

In the following analysis, we assign the phases of the correlated hoppings as 
\begin{align} \label{eq:FixedFlux}
\begin{split}
\calA_\parallel &= \pi-\alpha, \\
\calB_\parallel &= \beta,
\end{split}
\end{align}
for the hoppings inside the chains [for the definitions of $\calA_\parallel$ and $\calB_\parallel$, see Eq.~\eqref{eq:para}], and
\begin{align} \label{eq:FixedFlux2}
\begin{split}
\calA^{(j,\ell)}_{(j+1,\ell-1)} &=  \calA^{(j,\ell)}_{(j+1,\ell)} =\pi \ell, \\
\calB^{(j,\ell)}_{(j+1,\ell-1)} &= \calB^{(j,\ell)}_{(j+1,\ell)} =\pi \ell , 
\end{split}
\end{align}
for the hoppings between neighboring chains. 
This assignment of the phases is shown in Fig.~\ref{fig:FixedFlux}. 
The continuum limit is then obtained by 
\begin{align}
\begin{split}
a_0 \calA^{(j,\ell)}_{(j+1,\ell-1)} &= a_0 \calA^{(j,\ell)}_{(j+1,\ell)} \to \pi x, \\
a_0 \calB^{(j,\ell)}_{(j+1,\ell-1)} &= a_0 \calB^{(j,\ell)}_{(j+1,\ell)} \to \pi x. 
\end{split}
\end{align}

\section{Quantum Hall phases} \label{sec:StableQHS}

We are now ready to analyze Hamiltonian \eqref{eq:Ham} expressed in terms of the bosonic fields. 
 The detail can be found in Appendix~\ref{app:AlternativeHam}.
On extracting the vertex operators that govern the low-energy physics, we find that, as expected, the background gauge fluxes play an important role. 
Because of the uniform flux $\pi$ for each hexagonal plaquette (as shown in Fig.~\ref{fig:Model}), the exponential factor $e^{i\pi x/a_0}$ appears in the interaction Hamiltonian \eqref{eq:IntHamString}; terms involving this factor rapidly oscillate in $x$ and will vanish after the spatial integration.
In order to have nonvanishing terms in Eq.~\eqref{eq:IntHamString}, this factor must be canceled by another exponential factor $e^{i2n \pi \brho x}$ with $n \in \mathbb{Z}$. 
Thus we require certain commensurate conditions on the mean density $\brho$ under which quantum Hall states can arise. 
Furthermore, the staggered flux $\alpha$ or $\beta$ for each triangle, with the help of the frustrated structure of the triangular plaquettes in our model, selects the chirality of those quantum Hall states.

We now state the central results of this work. 
In the following, we consider the two cases with the filling factors $\brho a_0 =1/2$ and $1/6$. 
At $\brho a_0=1/2$, we obtain the BIQH phase with the electric Hall conductance $|\sigma_{xy}|=2$ as numerically found in Ref.~\cite{YCHe15a}. 
We also consider the effect of a perturbation that reduces the $U(1)_c \times U(1)_s$ symmetry to the $U(1)_c$ symmetry. 
Then we argue that the transition between the BIQH phases is described by two copies of the $O(4)$ nonlinear sigma model with $\theta=\pi$ when the Hamiltonian has time-reversal symmetry. 
At $\brho a_0 =1/6$, we expect a fractional quantum Hall state called the Halperin (221) state with $|\sigma_{xy}|=2/3$. 

\subsection{$\brho a_0 =1/2$: bosonic integer quantum Hall state} \label{sec:BIQHPhase}

At the filling $\brho a_0 = 1/2$, the nonvanishing terms in the interaction Hamiltonian are given by
\begin{align} \label{eq:SPTHam1}
H_1 \sim& -\frac{t' c_1^2}{2a_0} \sum_{j=1}^N \int dx \nonumber \\
& \times \Bigl[ (e^{-i\alpha}+e^{i\pi/2}) e^{i\left( \varphi^a_j +\theta^b_j +\pi N^b_j -\varphi^a_{j+1} +\theta^b_{j+1} \right)} \nonumber \\
& +(e^{-i\alpha}+e^{-i\pi/2}) e^{i\left( \varphi^a_j -\theta^b_j +\pi N^b_j -\varphi^a_{j+1} -\theta^b_{j+1} \right)} \nonumber \\
& +(e^{i\beta}+e^{-i\pi/2}) e^{i\left( \varphi^b_j +\theta^a_j +\pi N^a_j -\varphi^b_{j+1} +\theta^a_{j+1} \right)} \nonumber \\
& +(e^{i\beta}+e^{i\pi/2}) e^{i\left( \varphi^b_j -\theta^a_j +\pi N^a_j -\varphi^b_{j+1} -\theta^a_{j+1} \right)} +\textrm{h.c.} \Bigr]. 
\end{align} 
Here we have dropped vertex operators involving $e^{in\theta^s_j}$ with $n \geq 2$ since they will be irrelevant as we will discuss below. 
For usual lattice bosons only with hopping terms, the vertex operator takes the form $e^{i(\varphi^s_j-\varphi^s_{j+1})}$ and leads to the Bose condensation with $\langle e^{i\varphi^s_j} \rangle \neq 0$. 
In our case, as a result of the correlated hopping, those terms do not appear in the theory. 
Instead, we have the vertex operators of $\varphi^s_j \pm \theta^{s'}_j$ with $s \neq s'$, which appear to be the bound states of the boson in one sublattice and the vortex in the other sublattice. 
These particle-vortex bound states, which naturally emerge in our model, may play the role corresponding to the mutual composite bosons \cite{Senthil13}.

In a similar manner to Ref.~\cite{YMLu12}, we now introduce new bosonic fields by 
\begin{align} \label{eq:BIQHField}
\begin{split}
\phi^1_j(x) &= \varphi^a_j(x) +\theta^b_j(x) +\pi N^b_{j' \geq j}, \\
\phi^2_j(x) &= \varphi^b_j(x) +\theta^a_j(x) +\pi N^a_{j'<j}, \\
\tphi^1_j(x) &= \varphi^a_j(x) -\theta^b_j(x) +\pi N^b_{j' \geq j}, \\
\tphi^2_j(x) &= \varphi^b_j(x) -\theta^a_j(x) +\pi N^a_{j'<j}, 
\end{split}
\end{align}
where we have used the shorthand notations $N^s_{j' \geq j} \equiv \sum_{j' \geq j} N^s_{j'}$ and $N^s_{j'<j} \equiv \sum_{j'<j} N^s_{j'}$. 
These fields satisfy the commutation relations, 
\begin{align} \label{eq:BIQHFieldComm}
\begin{split}
[\partial_x \phi^\mu_j(x), \phi^\nu_{j'}(x')] &= 2i\pi \delta_{jj'} K_{\mu \nu} \delta(x-x'), \\
[\partial_x \tphi^\mu_j(x), \tphi^\nu_{j'}(x')] &= -2i\pi \delta_{jj'} K_{\mu \nu} \delta(x-x'),  \\
[\partial_x \phi^\mu_j(x), \tphi^\nu_{j'}(x')] &= 0,
\end{split}
\end{align}
where the matrix $\bfK$ is given by 
\begin{align} \label{eq:KMat}
\bfK = \left( \begin{array}{cc} 0 & 1 \\ 1 & 0 \end{array} \right). 
\end{align}
This $K$ matrix is nothing but the one appearing in the Chern-Simons theory of the BIQH state \cite{YMLu12}. 
The zero-mode operators $\pi N^s_j$ are necessary to be added in Eq.~\eqref{eq:BIQHField} so that any pair of the vertex operators commute with each other. 
This is not the case when we start from the array of \emph{ordinary} bosonic chains as in Ref.~\cite{YMLu12}. 
Since any vertex operator contains $e^{in\theta^s_j}$ with even $n$ in that case, the vertex operators obey commutation relations. 
In our case, because of the string operators, the vertex operators can be built from $e^{in\theta^s_j}$ with odd $n$ and therefore obeys anticommutation relations. 
This is compensated by the zero-mode operators $\pi N^s_j$, just like the Klein factors in fermionic systems. 

In terms of the new bosonic fields, the chain Hamiltonian \eqref{eq:ChainBoson} becomes 
\begin{align} \label{eq:ChainBoson2}
H_0 \sim \sum_{j=1}^N \sum_{\mu=1,2} \frac{v}{4\pi} \int dx \left[ (\partial_x \phi^\mu_j)^2 +(\partial_x \tphi^\mu_j)^2 \right]. 
\end{align}
Equation~\eqref{eq:SPTHam1} can be rewritten as 
\begin{align} \label{eq:SPTHam2}
H_1 \sim& -\frac{t'c_1^2}{2a_0} \sum_{j=1}^N \int dx \biggl[ g(\alpha) \cos \left( \phi^1_j -\tphi^1_{j+1} +\gamma(\alpha) \right) \nonumber \\
& +g(-\alpha) \cos \left( \tphi^1_j -\phi^1_{j+1} -\gamma(-\alpha) \right) \nonumber \\
& +g(\beta) \cos \left( \phi^2_j -\tphi^2_{j+1} -\gamma(\beta) \right) \nonumber \\
& +g(-\beta) \cos \left( \tphi^2_j -\phi^2_{j+1} +\gamma(-\beta) \right) \biggr], 
\end{align}
where $g$ and $\gamma$ are functions of the staggered fields given by
\begin{subequations}
\begin{align}
\label{eq:gandgamma}
g(\alpha) &= \sqrt{2-2\sin \alpha}, \\
\gamma(\alpha) &= \tan^{-1} \left[ \frac{1-\sin \alpha}{\cos \alpha} \right]. 
\end{align} 
\end{subequations}
As we can see from that the field $\phi^1_j -\tphi^1_{j+1}$ does not commute with $\tphi^2_{j-1} -\phi^2_j$ and $\tphi^2_{j+1}-\phi^2_{j+2}$, the two sets of the fields $\{ \phi^1_j-\tphi^1_{j+1} \}$ and $\{ \tphi^2_j -\phi^2_{j+1} \}$ cannot be localized simultaneously in minma of the cosine potentials. 
Similarly, $\{ \phi^2_j-\tphi^2_{j+1} \}$ and $\{ \tphi^1_j -\phi^1_{j+1} \}$ cannot be localized simultaneously. 
On the other hand, any pair of fields commute with each other within each of the two sets $\{ \phi^\mu_j -\tphi^\mu_{j+1} \}$ and $\{ \tphi^\mu_j -\phi^\mu_{j+1} \}$, so that any pair of fields from each of the two sets can be localized simultaneously and produce a gap in the bulk excitation spectrum. 

Since all the vertex operators have the same scaling dimension $1$ at the Gaussian fixed point \eqref{eq:ChainBoson2}, which corresponds to free fermions, only the relative strength of the coupling constants determines which term reaches the strong-coupling limit faster under the renormalization group transformation. 
The coupling constants obey 
\begin{align}
\begin{split}
\left| \frac{g(\alpha)}{g(-\alpha)} \right|>1 & \hspace{10pt} \textrm{for} \hspace{10pt} 0 < \alpha < \pi.
\end{split}
\end{align}
To be more specific, let us consider the case for $\alpha = \beta$. 
If $0 < \alpha < \pi$, $g(\alpha)$ goes the strong-coupling limit and all the fields in the bulk, $\{ \phi^\mu_j -\tphi^\mu_{j+1} \}$, acquire a gap. However, the unpaired edge modes $\tphi^\mu_1$ and $\phi^\mu_N$ remain gapless at the leftmost and rightmost chains. 
Similarly, for $-\pi < \alpha < 0$, $g(-\alpha)$ goes the strong-coupling limit and the edge modes $\phi^\mu_1$ and $\tphi^\mu_N$ remain gapless. 
The two BIQH phases are then characterized by the edge modes $\{ \tphi^\mu_1, \phi^\mu_N \}$ and $\{ \phi^\mu_1, \tphi^\mu_N \}$; we denote them as the BIQH${}^+$ and BIQH${}^-$ phases, respectively. 

In order to demonstrate that the two BIQH phases have opposite Hall conductances, let us consider how the bosonic fields transform under the $U(1)$ symmetries. 
As discussed earlier,  Hamiltonian \eqref{eq:OriginalHam} is invariant under the $U(1)_c \times U(1)_s$ symmetry, that is the symmetry under 
\begin{align} \label{eq:U1cU1s}
\begin{split}
a_{j,\ell} &\to e^{-i(\Delta \omega_c +\Delta \omega_s)} a_{j,\ell}, \\
b_{j,\ell} &\to e^{-i(\Delta \omega_c -\Delta \omega_s)} b_{j,\ell}, 
\end{split}
\end{align}
for arbitrary angles $\Delta \omega_c, \Delta \omega_s \in [0,2\pi)$. 
In the language of the bosonic fields, these transformation can be expressed as 
\begin{align} \label{eq:U1U1Sym}
\begin{split}
\varphi^a_j(x) &\to \varphi^a_j(x) + \Delta \omega_c +\Delta \omega_s, \\
\theta^a_j (x) &\to \theta^a_j(x), \\
\varphi^b_j(x) &\to \varphi^b_j(x) + \Delta \omega_c -\Delta \omega_s, \\
\theta^b_j(x) &\to \theta^b_j(x). 
\end{split}
\end{align}
Thus we have 
\begin{align} \label{eq:U1cU1sField}
\begin{split}
\phi^1_j(x) &\to \phi^1_j(x) +\Delta \omega_c +\Delta \omega_s, \\
\phi^2_j(x) &\to \phi^2_j(x) +\Delta \omega_c -\Delta \omega_s, \\
\tphi^1_j(x) &\to \tphi^1_j(x) +\Delta \omega_c +\Delta \omega_s, \\
\tphi^2_j(x) &\to \tphi^2_j(x) +\Delta \omega_c -\Delta \omega_s. 
\end{split}
\end{align}
For $\phi^\mu_j$, the basis diagonalizing the $K$ matrix in Eq.~\eqref{eq:BIQHFieldComm} is introduced by $\phi^1_j+\phi^2_j$ and $\phi^1_j-\phi^2_j$. 
From the above transformations, the former is the electric mode that carries only the charge $+2$, while the latter is the pseudospin mode that carries only the pseudospin $+2$. 
Since their corresponding eigenvalues of $K$ have different signs, those electric and pseudospin modes are counter-propagating at the edges $j=1$ and $N$.
Accordingly, the electric and pseudospin Hall conductances are quantized to $\sigma_{xy}=2$ and $\sigma^s_{xy}=-2$, respectively, resulting in a vanishing thermal Hall conductance \cite{YMLu12}. 
A similar discussion applies to $\tphi^\mu_j$ except for that the associated $K$ matrix has an overall minus sign. 
Hence, the edge currents of $\tphi^\mu_j$ flow in the opposite direction to those of $\phi^\mu_j$; the conductances become $\sigma_{xy}=-2$ and $\sigma^s_{xy}=2$. 
Therefore, the BIQH${}^+$ and BIQH${}^-$ phases have different signs in the Hall conductances.

At $\alpha=-\beta$, the vertex operators with $\{ \phi^\mu_j-\tphi^\mu_{j+1} \}$ and $\{ \tphi^\nu_j -\phi^\nu_{j+1} \}$ for $\mu \neq \nu$ have the same magnitude in the coupling constants. 
However, those fields are not commuting and thus cannot be simultaneously localized. 
We expect that this point describes the phase transition between the BIQH${}^+$ and BIQH${}^-$ phases. 

For general values of $\alpha$ and $\beta$, we obtain the phase diagram by using the renormalization group analysis (see Appendix~\ref{app:RG}), as shown in Fig.~\ref{fig:PhaseDiagram}. 
\begin{figure}
\includegraphics[clip,width=0.3\textwidth]{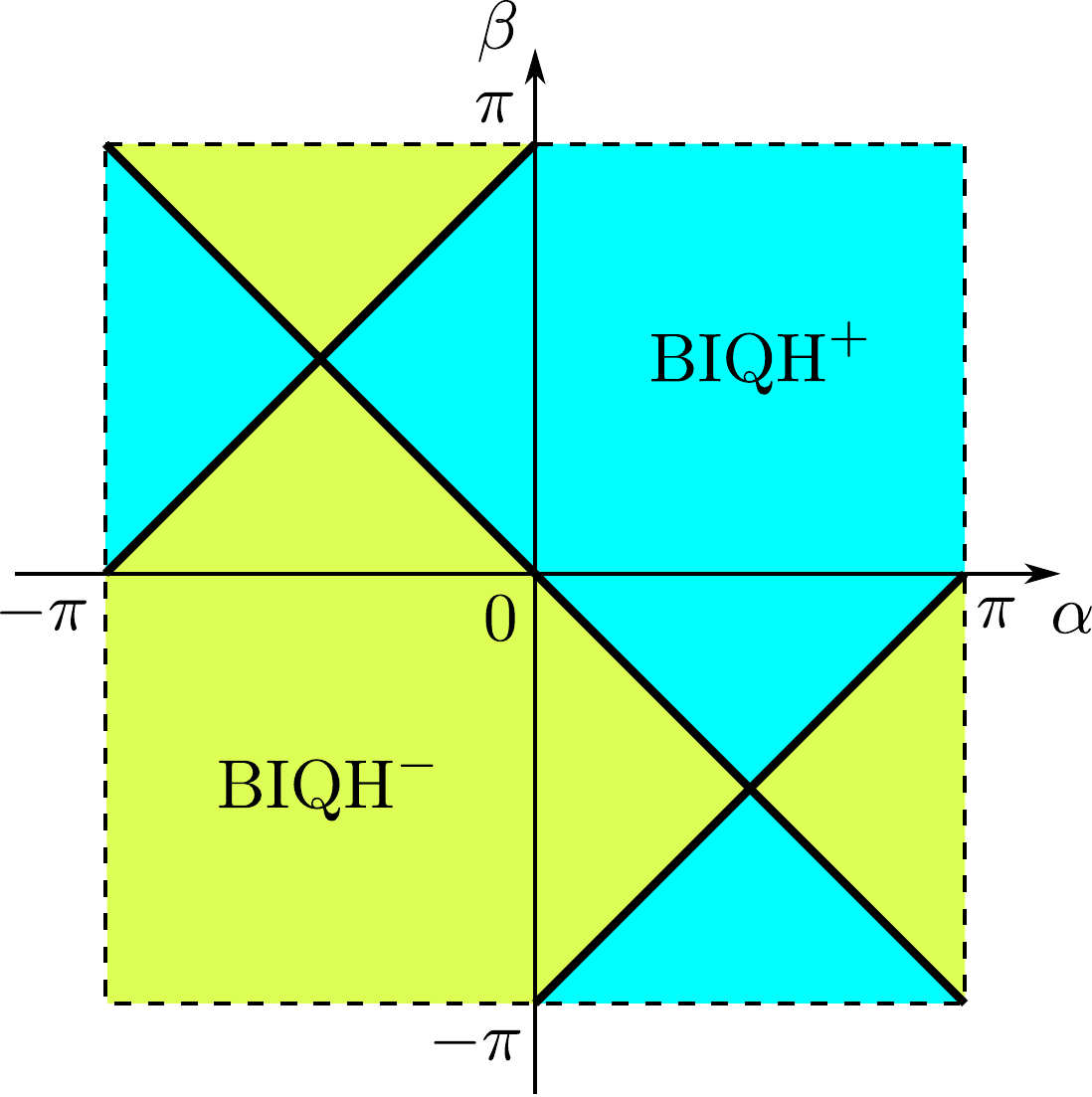}
\caption{(Color online) Phase diagram of the Hamiltonian \eqref{eq:Ham} for general values of $\alpha$ and $\beta$, obtained by the renormalization group analysis.
Two BIQH phases denoted by BIQH${}^+$ and BIQH${}^-$ (see text) are separated by phase transitions corresponding to black solid lines.}
\label{fig:PhaseDiagram}
\end{figure}
The phase transition is determined by the fixed points at which all the coupling constants flow to the same value; this occurs at the initial coupling $g(\alpha)=g(-\beta)$. 
Of course, the above phase diagram is valid only for the quasi-one-dimensional limit so that it is not clear whether this is applicable for the truly 2D model. 
Nevertheless, this phase diagram perfectly coincides with the one numerically obtained by infinite density-matrix renormalization group for the original Hamiltonian \eqref{eq:OriginalHam}. 
The nature of this transition is discussed in Sec.~\ref{sec:Transition}. 

We note that in Eq.~\eqref{eq:SPTHam2} we can in general add other vertex operators involving $e^{in\theta^s_j}$ with $n \geq 2$. 
These can potentially drive the system into (fractional) quantum Hall phases described by the $K$ matrix,
\begin{align} \label{eq:GeneralK}
\bfK_{(p,q)} = \left( \begin{array}{cc} 2p & 2q+1 \\ 2q+1 & 2p \end{array} \right), 
\end{align}
where $p$ and $q$ are integers. 
However, the corresponding vertex operator has the scaling dimension 
\begin{align}
x_{(p,q)}=\frac{4p^2 +(2q+1)^2+ 1}{2}.
\end{align}
Therefore, the BIQH phase is most promising at the Gaussian fixed point since the corresponding vertex operators ($p=q=0$) are strictly relevant. 
In fact, other vertex operators that drive the system to conventional ordered phases are at most marginal. 
Thus the BIQH phase is robust against any other competing order. 
This can be contrasted to the situation in Ref.~\cite{Gorohovsky15}, where the operators leading to the chiral spin liquid are marginal.

\subsection{Effect of tunneling between two sublattices} \label{sec:SBTerm}

\begin{figure}
\includegraphics[clip,width=0.4\textwidth]{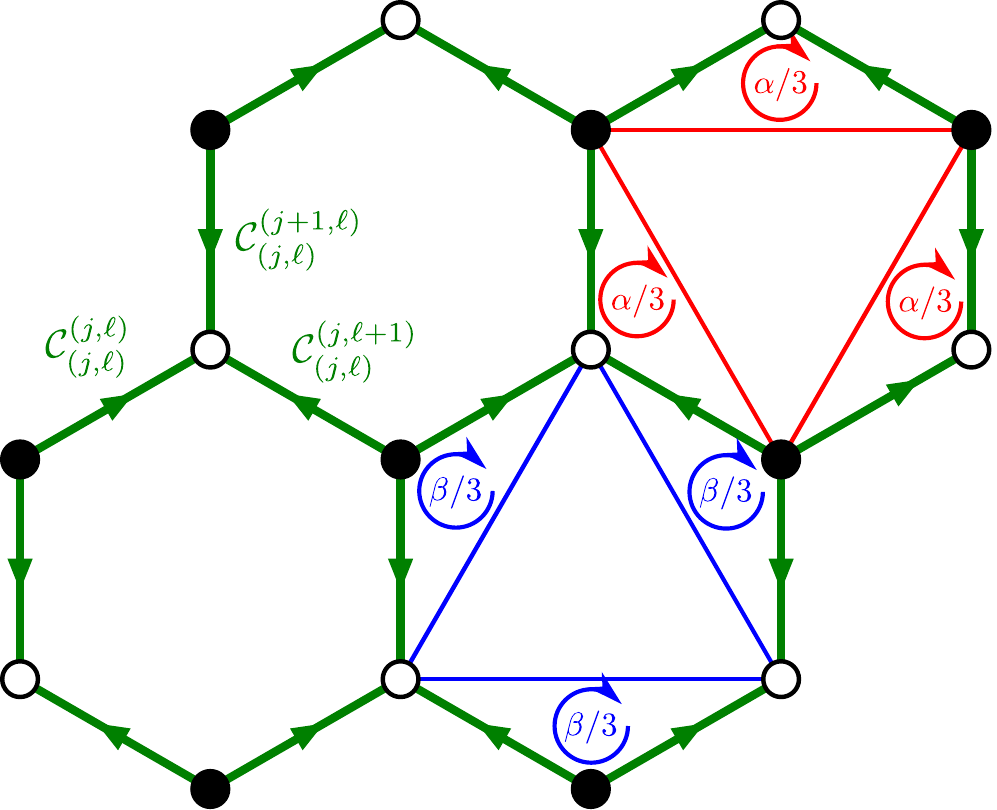}
\caption{(Color online) Tunnelings between the two sublattices are represented by green arrows on the links of the honeycomb lattice. 
We also show the background fluxes assigned in Ref.~\cite{YCHe15a}, where the fluxes $\alpha/3$ or $\beta/3$ thread obtuse triangles whose two vertices are in the $A$ or $B$ sublattice.}
\label{fig:SBTunneling}
\end{figure}

In the Hamiltonian given by Eq.~\eqref{eq:OriginalHam}, both the species of bosons are separately conserved. 
This gives rise to the $U(1)_a\times U(1)_b$ symmetry. 
However, upon introducing nearest-neighbor hoppings, which allow inter-species tunneling, only the global $U(1)$ charge corresponding to the conservation of the total Boson number survives. 
The BIQH phase described above is stable to such inter-species tunnelings (see Ref.~\cite{YCHe15a}), as we shall show now. 
To this end, we consider the following form of the tunnelings defined on the honeycomb lattice \cite{YCHe15a}, 
\begin{align} \label{eq:SBTunneling}
H_\lambda =& \ \lambda \sum_{j=1}^N \sum_{\ell=1}^L \Bigl[ e^{i\calC^{(j,\ell)}_{(j,\ell)}} a^\dagger_{j,\ell} b_{j,\ell} +e^{-i\calC^{(j,\ell+1)}_{(j,\ell)}} b^\dagger_{j,\ell} a_{j,\ell+1} \nonumber \\
& +e^{-i\calC^{(j+1,\ell)}_{(j,\ell)}} b^\dagger_{j,\ell} a_{j+1,\ell} +\textrm{h.c.} \Bigr]. 
\end{align}
These tunnelings are schematically shown in Fig.~\ref{fig:SBTunneling}. 

In the continuum limit, these tunnelings give rise to oscillating factors depending on the background fluxes. 
These oscillating factors must be canceled by certain commensurability conditions with the fluxes $\calC$ of the tunnelings. 
In general, this gives rise to the following \emph{intrachain} coupling terms 
\begin{align}
H^\parallel_\lambda \sim& \ \frac{\lambda c_1^2}{2a_0} \sum_{j=1}^N \int dx \nonumber \\
& \times \Bigl[ h^\parallel_1 \cos (\phi^1_j -\phi^2_j +\xi^\parallel_1) +h^\parallel_2 \cos (\tphi^1_j -\tphi^2_j +\xi^\parallel_2) \nonumber \\
&+h^\parallel_3 \cos (\phi^1_j-\tphi^2_j +\xi^\parallel_3) +h^\parallel_4 \cos (\tphi^1_j-\phi^2_j +\xi^\parallel_4) \Bigr]. 
\end{align}
Here $h^\parallel_n$ and $\xi^\parallel_n$ are some functions of $\alpha$, $\beta$, and $\calC$. 
At $\rho a_0 =1/2$ and under the background fluxes assigned in Ref.~\cite{YCHe15a} (see Fig.~\ref{fig:SBTunneling}), the first two terms appear when $(\alpha+\beta)/3=0$ mod $2\pi$ is satisfied, while the latter two terms appear when $(\alpha+\beta)/3=\pi$ mod $2\pi$ is satisfied. 
If $\lambda$ is sufficiently smaller than the interchain coupling $t'$, those interactions may be neglected in the bulk. 
However, at the edges $j=1$ and $j=N$, the first two terms couple with the pseudospin modes $\phi^1_j -\phi^2_j$ and $\tphi^1_j -\tphi^2_j$. 
Although at first sight they appear to open gaps for the pseudospin modes at the edges, those tunnelings cannot localize the pseudospin modes since they do not satisfy the Haldane's null-vector criterion \cite{Haldane95}; indeed each of them does not commute with itself since they are chiral.
Thus the pseudospin edge modes still remain gapless even after introducing the tunneling between the two sublattices. 

From the point of view of the symmetry, this can be easily understood as follows: 
Without the tunneling \eqref{eq:SBTunneling}, the system maintains the $U(1)_a \times U(1)_b$ symmetry, or equivalently the $U(1)_c \times U(1)_s$ symmetry defined in Eq.~\eqref{eq:U1cU1s}. 
From Eq.~\eqref{eq:U1cU1sField}, this symmetry forbids any perturbation of the form $\cos(n\phi^\mu_j)$ or $\cos(n\tphi^\mu_j)$ with integer $n$, which gaps out the edge modes. 
Thus we have quantized electric and pseudospin Hall conductances in the absence of the tunneling \eqref{eq:SBTunneling}. 
However, once the tunneling is turned on, the $U(1)_c \times U(1)_s$ symmetry reduces to the $U(1)_c$ symmetry associated with the global particle conservation.
Therefore the perturbations of the form $\cos[n(\phi^1_j-\phi^2_j)]$ and $\cos[n(\tphi^1_j-\tphi^2_j)]$ are allowed. 
However, those interactions must carry nonzero conformal spins and cannot satisfy the Haldane's null-vector criterion \cite{Haldane95,Mulligan14}. 
Thus the counter-propagating gapless edge modes cannot be gapped as far as we keep the global particle conservation. 
This can also be understood that the $U(1)_c$ symmetry ``chirally'' acts on the edge theory \cite{XChen12b};
the $U(1)_c$ symmetry acts on the left- and right-moving modes in different ways and therefore forbids the back-scattering gapping out the edge modes. 

Aside from the intrachain couplings, there are also \emph{interchain} couplings given by 
\begin{align}
H^\perp_\lambda \sim& \ \frac{\lambda c_1^2}{2a_0} \sum_{j=1}^N \int dx \Bigl[ h^\perp_1 \cos(\phi^2_j-\tphi^1_{j+1} +\xi^\perp_1) \nonumber \\
&+h^\perp_2 \cos(\tphi^2_j-\phi^1_{j+1}+\xi^\perp_2) \Bigr] \nonumber \\
&+\frac{\lambda'}{a_0} \sum_{j=1}^N \int dx \Bigl[ h^\perp_3 \cos(\phi^1_j-\tphi^2_{j+1}+\xi^\perp_3) \nonumber \\
&+h^\perp_4 \cos(\tphi^1_j-\phi^2_{j+1} +\xi^\perp_4) \Bigr], 
\end{align}
where $h^\perp_n$ and $\xi^\perp_n$ are some functions of $\alpha$, $\beta$, and $\calC$. 
Although the $\lambda'$ terms are not included in the bare Hamiltonian, they can be generated by higher-order perturbations. 
Since all the fields appearing in the cosine potentials do not commute with any of those in Eq.~\eqref{eq:SPTHam2}, the tunneling does not modify the qualitative nature of the BIQH phases as long as $\lambda$ is sufficiently smaller than $t'$. 
Although the tunnelings do not alter the physics deeply inside the BIQH phase, their effect may be crucial in the vicinity of the transition between the BIQH phases as we discuss below. 

\subsection{Transition between BIQH phases} \label{sec:Transition}

We here discuss the nature of the phase transition between the two BIQH phases with the Hall conductances $\sigma_{xy}= \pm 2$. 
Indeed, if a direct continuous transition is possible, it falls beyond traditional Landau-Ginzburg-Wilson paradigm since none of the two phases break any symmetry spontaneously and therefore are not described in terms of local order parameter fields. 
We discuss that this transition can be described by two copies of the $O(4)$ nonlinear sigma model (NLSM) with the topological $\theta$ term and thus is possibly in a deconfined quantum criticality \cite{Senthil04a,Senthil04b}.

We below proceed in the following steps: 
We first show that the chain Hamiltonian \eqref{eq:ChainBoson2} is equivalent to the $SU(2)_1$ Wess-Zumino-Witten (WZW) models and then the interaction \eqref{eq:SPTHam2} can be expressed in terms of $SU(2)$-matrix fields. 
It reveals that the Hamiltonian possesses an $SO(4) \times SO(4)$ symmetry for special values of $\alpha$ and $\beta$. 
Following the argument by Senthil and Fisher \cite{Senthil06}, we show that the Hamiltonian is equivalent to two copies of the $(2+1)$-dimensional $O(4)$ NLSM with $\theta=\pi$, which are presumably equivalent to two copies of the two-flavor massless QED${}_3$. 

Let us introduce chiral bosonic fields corresponding to charge and pseudospin as
\begin{align}
\begin{split}
\phi^1_j(x) &= \phi^R_{c,j}(x)-\phi^L_{s,j}(x), \\
\phi^2_j(x) &= \phi^R_{c,j}(x)+\phi^L_{s,j}(x), \\
\tphi^1_j(x) &= -\phi^R_{s,j}(x)+\phi^L_{c,j}(x), \\
\tphi^2_j(x) &= \phi^R_{s,j}(x)+\phi^L_{c,j}(x). 
\end{split}
\end{align}
These fields satisfy the commutation relations, 
\begin{align}
\begin{split}
[\partial_x \phi^R_{\rho,j}(x), \phi^R_{\rho',j'}(x')] &= i\pi \delta_{\rho \rho'}\delta_{jj'} \delta(x-x'), \\
[\partial_x \phi^L_{\rho,j}(x), \phi^L_{\rho',j'}(x')] &= -i\pi \delta_{\rho \rho'}\delta_{jj'} \delta(x-x'), 
\end{split}
\end{align}
with $\rho = c,s$. 
At the boundary $j=1$ or $N$, those fields represent nothing but the counter-propagating gapless edge modes discussed before. 

Each chain Hamiltonian is composed of the two hard-core bosonic chains. 
As it can be mapped onto the tight-binding Hamiltonian of spinful electrons, it possesses a charge $SU(2)$ symmetry as well as a pseudospin $SU(2)$ symmetry at half filling. 
Thus the chain Hamiltonian \eqref{eq:ChainBoson2} is described by $2N$ copies of the $SU(2)_1$ WZW theory \cite{Affleck86,Affleck87}, 
\begin{align} \label{eq:ChainWZW}
H_0 \sim \frac{v}{6\pi} \sum_{j=1}^N \sum_{\rho=c,s} \int dx \left[ :\mathrel{\bfJ^R_{\rho,j} \cdot \bfJ^R_{\rho,j}}: + :\mathrel{\bfJ^L_{\rho,j} \cdot \bfJ^L_{\rho,j}}: \right], 
\end{align}
where $\bfJ^{R/L}_{\rho,j}$ are the $SU(2)_1$ currents related to the bosonic fields as 
\begin{align}
(\bfJ^{R/L}_{\rho,j})^z = a_0 \partial_x \phi^{R/L}_{\rho,j}, \hspace{10pt} (\bfJ^{R/L}_{\rho,j})^\pm = e^{\pm i2 \phi^{R/L}_{\rho,j}}, 
\end{align}
and $:\mathrel{X}:$ means the normal-ordered product of $X$. 
Then we introduce the $SU(2)$-matrix fields \cite{Tsvelik}, 
\begin{align}
\begin{split}
(\bfg_j)_{\sigma \sigma'} &= \frac{e^{i\pi/4}}{\sqrt{2}} (\bfz^L_{s,j})_\sigma (\bfz^{R \dagger}_{c,j})_{\sigma'}, \\
(\tbfg_j)_{\sigma \sigma'} &= \frac{e^{i\pi/4}}{\sqrt{2}} (\bfz^{L \dagger}_{c,j})_\sigma (\bfz^R_{s,j})_{\sigma'}, 
\end{split}
\end{align}
through the spinor fields, 
\begin{align}
\bfz^{R/L}_{\rho,j} = \left( \begin{array}{c} e^{i\phi^{R/L}_{\rho,j}} \\ e^{-i\phi^{R/L}_{\rho,j}} \end{array} \right). 
\end{align}
Those spinor fields correspond to two primary fields of the $SU(2)_1$ WZW theory in each chiral sector. 
Specifically, these matrix fields are given by 
\begin{align}
\begin{split}
\bfg_j &= \frac{1}{\sqrt{2}} \left( \begin{array}{cc} e^{-i\phi^1_j} & ie^{i\phi^2_j} \\ ie^{-i\phi^2_j} & e^{i\phi^1_j} \end{array} \right), \\
\tbfg_j &= \frac{1}{\sqrt{2}} \left( \begin{array}{cc} e^{-i\tphi^1_j} & ie^{i\tphi^2_j} \\ ie^{-i\tphi^2_j} & e^{i\tphi^1_j} \end{array} \right). 
\end{split}
\end{align}

From Eq.~\eqref{eq:ChainWZW}, it is now transparent that in each of charge and pseudospin sectors, the chain Hamiltonian has an $SU(2)_R \times SU(2)_L$ symmetry. 
To be more precise, the $SU(2)_{c,R} \times SU(2)_{s,L}$ symmetry is endowed by the fields $\phi^\mu_j$, while the $SU(2)_{s,R} \times SU(2)_{c,L}$ symmetry is endowed by the fields $\tphi^\mu_j$. 
These give two different realizations of the $SO(4) \sim SU(2)_R \times SU(2)_L$ symmetry in each chain. 
We below mention how these $SU(2)$ symmetries act on the $SU(2)$-matrix fields. 
For the charge sector, the $SU(2)_{c,R}$ symmetry acts on the field $\bfg_j$ as a left multiplication of $\bfU \in SU(2)$, $\bfg_j \to \bfg_j \bfU$, while the $SU(2)_{c,L}$ symmetry acts on $\tbfg_j$ as $\tbfg_j \to \tbfg_j \bfV$ with $\bfV \in SU(2)$. 
For the pseudospin sector, the $SU(2)_{s,R}$ and $SU(2)_{s,L}$ symmetries respectively act on $\tbfg_j$ and $\bfg_j$ as the right multiplications of elements of $SU(2)$. 
This can be easily seen from the fact that the charge $U(1)$ symmetry given by Eq.~\eqref{eq:U1U1Sym} acts as $\bfg_j \to \bfg_j e^{-i\Delta \omega_c \sigma^z}$ and $\tbfg_j \to \tbfg_j e^{-i\Delta \omega_c \sigma^z}$, where $\sigma^{x,y,z}$ is the Pauli matrices. 
The pseudospin $U(1)$ symmetry acts as $\bfg_j \to e^{-i\Delta \omega_s \sigma^z} \bfg_j$ and $\tbfg_j \to e^{-i\Delta \omega_s \sigma^z} \tbfg_j$. 

We can now rewrite the interaction Hamiltonian \eqref{eq:SPTHam2} in terms of the matrix fields. 
For example, the first and third terms of Eq.~\eqref{eq:SPTHam2} are expressed as 
\begin{subequations} \label{eq:GMatCos}
\begin{align}
& g(\alpha) \cos \left( \phi^1_j-\tphi^1_{j+1}+\gamma(\alpha) \right) \nonumber \\
&= \frac{g(\alpha) \cos \gamma(\alpha)}{4} \textrm{Tr} (\bfg^\dagger_j \tbfg_{j+1} +\bfg^\dagger_j \sigma^z \tbfg_{j+1} \sigma^z) \nonumber \\
&+\frac{ig(\alpha) \sin \gamma(\alpha)}{4} \textrm{Tr} (\bfg^\dagger_j \tbfg_{j+1} \sigma^z +\bfg^\dagger_j \sigma^z \tbfg_{j+1}) +\textrm{h.c.}, \\
& g(\beta) \cos \left( \phi^2_j -\tphi^2_{j+1} -\gamma(\beta) \right) \nonumber \\
&= \frac{g(\beta) \cos \gamma(\beta)}{4} \textrm{Tr} (\bfg^\dagger_j \tbfg_{j+1} -\bfg^\dagger_j \sigma^z \tbfg_{j+1} \sigma^z) \nonumber \\
&-\frac{ig(\beta) \sin \gamma(\beta)}{4} \textrm{Tr} (\bfg^\dagger_j \tbfg_{j+1} \sigma^z -\bfg^\dagger_j \sigma^z \tbfg_{j+1}) +\textrm{h.c.}. 
\end{align}
\end{subequations}
Thus we may write 
\begin{align}
H_1 \sim H_{SO(4)} +H_{U(1) \times U(1)}, 
\end{align}
where the first term is the $SO(4)$ symmetric part,
\begin{align}
H_{SO(4)} = -\sum_{j=1}^N \int dx \ \textrm{Tr}(t_1 \bfg^\dagger_j \tbfg_{j+1} +t_2 \tbfg^\dagger_j \bfg_{j+1}) +\textrm{h.c.}, 
\end{align}
and the second term represents anisotropies that reduce the $SO(4)$ symmetry to the $U(1) \times U(1)$ symmetry, which is the genuine microscopic symmetry in the absence of inter-species tunneling. 
As seen from Eq.~\eqref{eq:GMatCos}, for general values of $\alpha$ and $\beta$, there exist the $U(1) \times U(1)$ anisotropies. 
However, at the time-reversal-symmetric points where $\alpha$ and $\beta$ take $0$ or $\pi$, the effective low-energy theory has the $SO(4)$ symmetry. 
In those cases, Eq.~\eqref{eq:SPTHam2} is invariant under the time-reversal symmetry $\phi^\mu_j \leftrightarrow -\tphi^\mu_j$ since $g(\alpha)=g(-\alpha)$ and $\gamma(\alpha)=\gamma(-\alpha)$ (the same relations hold for $\beta$). 
We further have $g(\alpha)=g(\beta)$ at those points, because of lattice symmetries that interchange the two sublattices. 
Then we can eliminate the phase shifts $\gamma$ by the redefinition of the fields, e.g. $\phi^1_j \to \phi^1_j-\gamma(\alpha)$ and $\phi^2_j \to \phi^2_j+\gamma(\beta)$. 

\begin{figure}
\includegraphics[clip,width=0.4\textwidth]{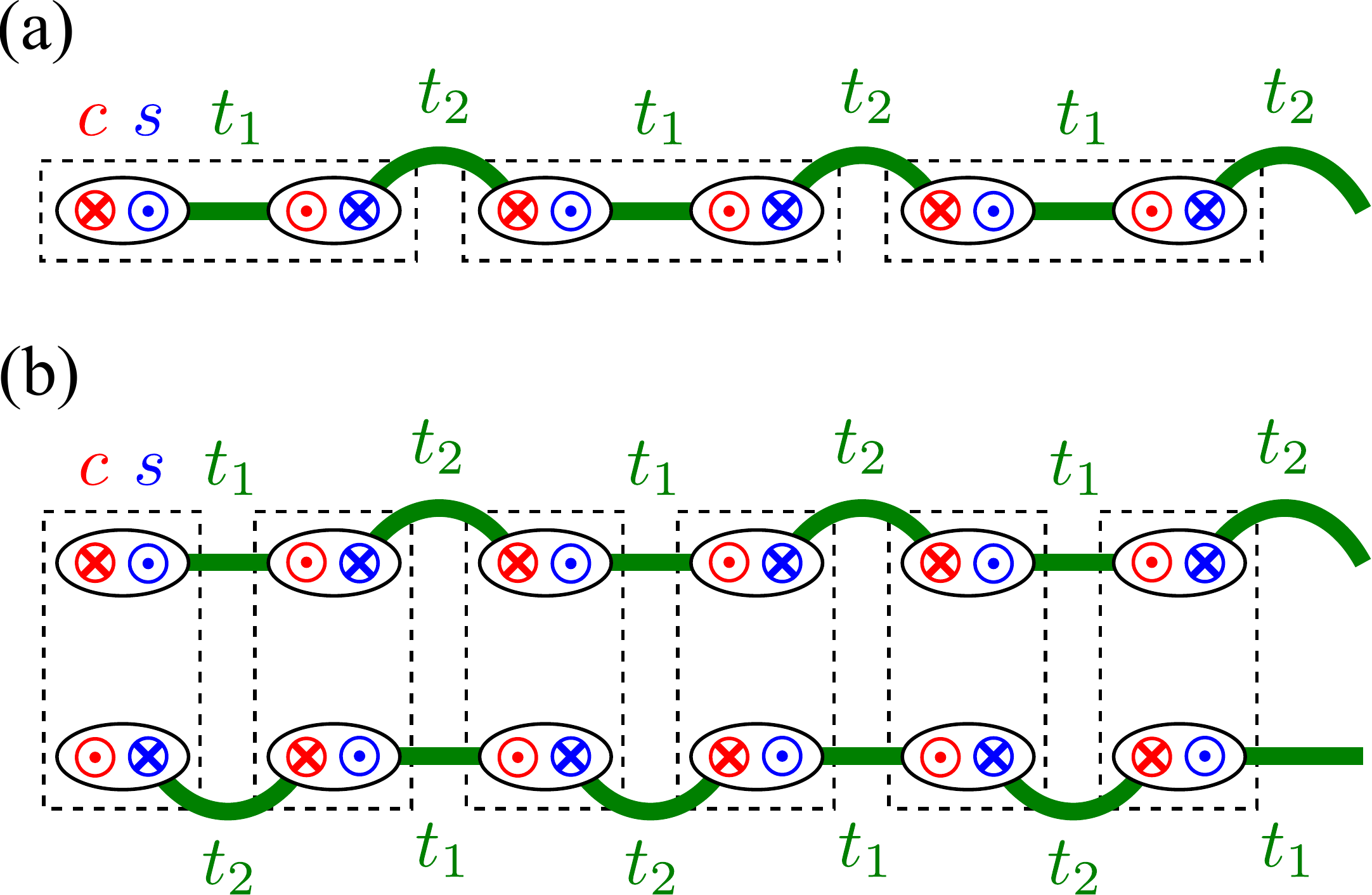}
\caption{(Color online) Network models describing the phase transitions between two BIQH phases. 
Red (blue) circles correspond to charge (pseudospin) modes, and cross (dotted) symbols represent right-moving (left-moving) modes. 
Four modes enclosed by the dashed line represent the degrees of freedom in a physical bosonic chain. 
$t_1$ and $t_2$ are two different tunnelings between the two counter-propagating modes. 
(a) The model studied in Refs.~\cite{Vishwanath13,XChen13b}: One may find the trivial phase when $t_1 \gg t_2$ while the BIQH${}^+$ phase when $t_1 \ll t_2$. 
(b) Our model: One may find the BIQH${}^-$ phase when $t_1 \gg t_2$ while the BIQH${}^+$ phase when $t_1 \ll t_2$. 
In both cases, the phase transitions between the two phases are expected at $t_1=t_2$.} 
\label{fig:NetworkModel}
\end{figure}

Thus, when the time-reversal symmetry exists, the phase transition between the BIQH${}^+$ and BIQH${}^-$ phases may be governed by the $SO(4)$ symmetric Hamiltonian $H_0+H_{SO(4)}$ with $t_1=t_2$. 
This Hamiltonian is similar to the network model studied for the phase transition between the BIQH and trivial Mott-insulating phases in Refs.~\cite{Vishwanath13,XChen13b}. 
The latter network model has an $SO(4)$ symmetry and is depicted in Fig.~\ref{fig:NetworkModel}~(a). 
However, our model takes a bilayer structure as shown in Fig.~\ref{fig:NetworkModel}~(b) and has an $SO(4) \times SO(4)$ symmetry. 
With $n$ being integer, one $SO(4)$ symmetry acts only on $\bfg_{2n}$ and $\tbfg_{2n+1}$, while the other acts only on $\bfg_{2n+1}$ and $\tbfg_{2n}$.
As discussed in Ref.~\cite{Senthil06} (see also Ref.~\cite{Vishwanath13}), after taking the continuum limit in the direction perpendicular to the chains, each layer of the network model gives the $(2+1)$-dimensional $O(4)$ NLSM with the topological $\theta$ terms at $\theta=\pi$. 
The action corresponding to our model is given by 
\begin{align}
S =& \ \frac{1}{g_0} \int d^3x \left[ \textrm{Tr}(\partial_\mu \bfG_1^\dagger \partial_\mu \bfG_1) +\textrm{Tr} (\partial_\mu \bfG_2^\dagger \partial_\mu \bfG_2) \right] \nonumber \\
&+i\pi S_\theta[\bfG_1] -i\pi S_\theta[\bfG_2], 
\end{align}
where $\bfG_{1,2}$ are the $SU(2)$-matrix field defined on the three-dimensional sphere $S^3$ and the $\theta$ term is given by
\begin{align}
S_\theta[\bfG] = \frac{i}{24\pi} \int d^3x \ \epsilon_{\mu \nu \rho} \textrm{Tr} (\bfG^\dagger \partial_\mu \bfG \bfG^\dagger \partial_\nu \bfG \bfG^\dagger \partial_\rho \bfG). 
\end{align}
Different signs in the $\theta$ terms for $\bfG_{1,2}$ originate from the fact that one layer of the network model has the opposite current structure to another layer [see Fig.~\ref{fig:NetworkModel} (b)].
Since the (2+1)-dimensional $O(4)$ NLSM at $\theta=\pi$ has been suggested to be equivalent to the two-flavor massless QED${}_3$ \cite{Senthil06,CXu15}, we have two copies of the two-flavor massless QED${}_3$ at the time-reversal-symmetric transition point between the two BIQH phases; two flavors of the Dirac fermion transform to another two under the time-reversal symmetry.
Possible transitions described by QED${}_3$ between BIQH phases have also been argued by the parton construction of the BIQH phase \cite{Grover13,YMLu14b}.

The BIQH phases can be obtained from the above NLSM as follows. 
Away from the critical point $t_1 = t_2$, a staggered structure of the tunneling may modify the coefficients of the $\theta$ terms as \cite{Tanaka06}
\begin{align}
i\pi (1-\delta) S_\theta[\bfG_1] -i\pi (1+\delta) S_\theta[\bfG_2], 
\end{align}
where we have set $t_1=t(1+\delta)$ and $t_2=t(1-\delta)$. 
For $t_1 \ll t_2$, the low-energy physics may be governed by the $O(4)$ NLSM with $\theta=2\pi$ for $\bfG_1$ and that with $\theta=0$ for $\bfG_2$. 
For a large value of $g_0$, the $O(4)$ NLSM is disordered in the bulk but has the critical (1+1)-dimensional $SU(2)_1$ WZW theories at the boundaries when $\theta=2\pi$ \cite{Senthil13,CXu13}. 
For $t_1 \gg t_2$, the physics may be governed by the $O(4)$ NLSM with $\theta=0$ for $\bfG_1$ and that with $\theta=-2\pi$ for $\bfG_2$. 
The $O(4)$ NLSM with $\theta=-2\pi$ also gives the boundary WZW theories but with the charge and pseudospin currents flowing in the opposite directions to those at $\theta=2\pi$. 

However, the BIQH phases break the time-reversal symmetry so that there always exist $U(1) \times U(1)$ anisotropies. 
Thus the $SO(4)$ symmetry must be reduced to the $U(1) \times U(1)$ symmetry. 
This is also the case for the phase transition between the BIQH phases when the time-reversal symmetry is broken. 
The resulting effective theory at the transition may be described by two copies of the easy-plane non-compact $CP^1$ model \cite{Senthil06}. 
The tunneling between the two sublattices further reduces the symmetry to a $U(1) \times Z_2$ symmetry. 
The fate of the transition may become (i) the first-order transition or (ii) some intermediate phase. 

\subsection{$\brho a_0 =1/6$: Halperin (221) state}

At the commensurate density $\brho a_0 =1/6 $, the vertex operators in Eq.~\eqref{eq:SPTHam2} vanish due to rapidly oscillating factors. 
Instead, the vertex operators corresponding to $p=1$ and $q=0$ in Eq.~\eqref{eq:GeneralK} are allowed. 
Then we have the interaction Hamiltonian,
\begin{align} \label{eq:221Ham}
H_1 \sim& -\frac{t' c_2^2}{8a_0} \sum_{j=1}^N \int dx \Bigl[ f(\alpha) \cos \left( \chi^1_j -\tchi^1_{j+1} +\eta(\alpha) \right) \nonumber \\
& +f(-\alpha) \cos \left( \tchi^1_j -\chi^1_{j+1} -\eta(-\alpha) \right) \nonumber \\
& +f(\beta) \cos \left( \chi^2_j -\tchi^2_{j+1} -\eta(\beta) \right) \nonumber \\
& +f(-\beta) \cos \left( \tchi^2_j -\chi^2_{j+1} +\eta(-\beta) \right) \Bigr], 
\end{align}
where we have defined the new bosonic fields, 
\begin{align} \label{eq:221Field}
\begin{split}
\chi^1_j(x) &= \varphi^a_j(x) +2\theta^a_j(x) +\theta^b_j(x) +\pi N^b_{j' \geq j}, \\
\chi^2_j(x) &= \varphi^b_j(x) +2\theta^b_j(x) +\theta^a_j(x) +\pi N^a_{j'<j}, \\
\tchi^1_j(x) &= \varphi^a_j(x) -2\theta^a_j(x) -\theta^b_j(x) +\pi N^b_{j' \geq j}, \\
\tchi^2_j(x) &= \varphi^b_j(x) -2\theta^b_j(x) -\theta^a_j(x) +\pi N^a_{j'<j}, 
\end{split}
\end{align}
and the functions $f$ and $\eta$ are given by 
\begin{subequations}
\begin{align}
f(\alpha) =& \ \sqrt{2+2\cos(\alpha-\pi/6)}, \\
\eta(\alpha) =& \ \tan^{-1} \left[ \frac{\cos(\alpha+\pi/3)+\sqrt{3}/2}{\sin(\alpha+\pi/3)+1/2} \right] \nonumber \\
& +\pi \left[ \Theta(-\alpha-\pi/2) +\Theta(-\alpha-5\pi/6) \right]. 
\end{align}
\end{subequations}
The new fields satisfy the commutation relations, 
\begin{align} \label{eq:221FieldComm}
\begin{split}
[\partial_x \chi^\mu_j(x), \chi^\nu_{j'}(x')] &= 2i\pi \delta_{jj'} K_{\mu \nu} \delta(x-x'), \\
[\partial_x \tchi^\mu_j(x), \tchi^\nu_{j'}(x')] &= -2i\pi \delta_{jj'} K_{\mu \nu} \delta(x-x'), \\
[\partial_x \chi^\mu_j(x), \tchi^\nu_{j'}(x')] &= 0, 
\end{split}
\end{align}
with the $K$ matrix, 
\begin{align} \label{eq:Kmat221}
\bfK = \left( \begin{array}{cc} 2 & 1 \\ 1 & 2 \end{array} \right).
\end{align}

The $K$ matrix \eqref{eq:Kmat221} indicates that the resulting gapped phase is in the Halperin $(221)$ state \cite{Halperin83}, which is a fractional quantum Hall state with the fractional electric Hall conductance $\sigma_{xy}=\pm 2/3$ and the integer pseudospin Hall conductance $\sigma_{xy}^s= \pm 2$. 
As opposed to the BIQH phase at $\brho a_0 = 1/2$, this phase supports chiral edge states, that is, the electric and pseudospin currents flow in the same direction. 
Moreover, the ground state on a torus is three-fold degenerate because of the anyonic statistics of the quasiparticle excitations. 
Thus we have the possibility of a topologically ordered phase in the same model but just at the different filling $\brho a_0=1/6$. 

As in the same manner of Sec.~\ref{sec:BIQHPhase}, the relative strength of the coupling constants $f(\alpha)$ and $f(-\alpha)$ determines the chirality of currents. 
However, the corresponding vertex operators now have the scaling dimension $3$ and thus irrelevant at the Gaussian fixed point. 
This suggests that for a perturbatively small $t'$, the ground state is in a sliding-Luttinger-liquid phase with forward scatterings generated by higher-order perturbations \cite{Vishwanath01} or some conventional ordered phase. 
Nevertheless, it is natural to expect that the Halperin $(221)$ state is realized for a certain strong value of $t'$ beyond the perturbative regime unless some competing ordered phase is found. 
This phase is also expected to be realized at the filling $\brho a_0 = 5/6$. 

Since our model favors the mutual composite boson, which is the bound state of a boson in one sublattice and a vortex in the other sublattice, the Halperin $(221)$ is also a natural candidate; 
the Halperin $(221)$ state is obtained by further attaching two flux quanta of itself to the mutual composite boson. 
Both the BIQH and Halperin $(221)$ states are in fact found in the two-component Bose gas with two-body interactions \cite{Furukawa13,YHWu13,Grass14}. 
We note that both of them have also been proposed to be realized in a spin-1 lattice Hamiltonian \cite{YMLu14a}. 
In that case, which state actually realizes depends on the Chern numbers of a free-fermion Hamiltonian before the projection onto the spin-1 Hilbert space. 
In our lattice model, both states can be potentially realized solely by tuning the filling factor of the bosons in the same Hamiltonian. 

\section{Conclusion} \label{sec:Conclusion}

We studied hard-core bosons interacting via correlated hoppings in an anisotropic deformation of the honeycomb lattice. 
A variant of the Jordan-Wigner transformation is introduced to resolve the correlated hopping in each chain, while it leaves nonlocal interactions between chains. 
Applying  bosonization techniques and using the the coupled-wire construction, we found that the BIQH phase is realized at half filling. 
Interestingly, in terms of the bosonic fields, the nonlocal interaction between chains can be seen as a hopping of the mutual composite bosons, which are bound states of the boson in one sublattice and the vortex in the other sublattice; by condensing them, one can obtain the BIQH phase \cite{Senthil13}. 
Thus it appears that the correlated hopping is a natural way to implement a mutual flux attachment in lattice systems.
We also discussed the stability of the BIQH phase against the tunneling between the two sublattices. 
Based on the effective $O(4)$ NLSM description, we argued that the transition between the two BIQH phases is in a deconfined quantum criticality when time-reversal symmetry is maintained. 
We further provided the possibility to stabilize the Halperin $(221)$ state at $1/6$ and $5/6$ fillings. 

Compared to  previous numerical results \cite{YCHe15a}, this work strongly suggests that the spatially anisotropic (quasi-one-dimensional) limit and the isotropic honeycomb lattice are smoothly connected for the BIQH state in the present model. 
Furthermore, the BIQH phases are found even when the bosons on the different sublattices feel different staggered fluxes.
On the other hand, a sliding Luttinger liquid will take the place of the Halperin $(221)$ state in the anisotropic limit since the corresponding interactions are irrelevant. 
However, as far as competing ordinarily ordered states are suppressed by frustration, the Halperin $(221)$ state is considered to be the most promising candidate of possible phases on the isotropic model. 
We leave the question about the actual realization of this state in the isotropic model for future work. 

In the present analysis, we obtained the effective description of the transition between the two BIQH phases in a somewhat intricate way; 
we first mapped our model to the $O(4)$ NLSM and then used the putative equivalence between the $O(4)$ NLSM with the $\theta$ term at $\theta=\pi$ and the two-flavor massless QED${}_3$. 
On the other hand, Mross, Alicea, and Motrunich recently demonstrated that the two-flavor QED${}_3$ can be directly obtained within the coupled-wire approach without passing through the NLSM \cite{Mross15b}. 
This alternative approach may provide more physical insights about the nature of the transition, e.g., the transformation properties of Dirac fermions under the symmetry. 
We also note that the model studied in this paper is in fact related to an effective lattice-gauge-theoretical description of the kagome antiferromagnet \cite{YCHe15b,YCHe15c}. 
Therefore a more detailed study of the transition will be an important task to understand the spin-liquid ground state of the kagome antiferromagnet. 

Needless to say, the present coupled-wire approach is not restricted to our specific example. 
The bosonization approach combined with the idea of coupled-wire construction to an anisotropic lattice will be applied to the broad class of interacting lattice models and topological phases. 
To propose experimentally relevant models for condensed matter or optical lattice systems in this way will be a more challenging but very interesting direction in the future. 

\section*{acknowledgements}
The authors acknowledge T. Meng for useful discussions and R. Moessner for discussions and previous collaboration on a related topic. 
Y.F. is grateful to S. Furukawa and P. Lecheminant for discussions and a related collaboration. 
S.B. acknowledges the visitors program at MPIPKS for hospitality. 

\appendix

\section{Alternative representation of the Hamiltonian} \label{app:AlternativeHam}

We here show that we can obtain the qualitatively same result when we start from the interaction Hamiltonian \eqref{eq:IntHamString} \emph{without} incorporating the density operators into the string operators. 
Let us first recall how Eq.~\eqref{eq:IntHamString} is bosonized, in which case the density operators have already been absorbed, by using Eqs.~\eqref{eq:DensityBoson} and \eqref{eq:StringBoson}. 
Focusing on the first two terms, in the continuum limit, the hopping perts are bosonized as 
\begin{subequations} \label{eq:HopBoson}
\begin{align}
&e^{i\calA^{(j,\ell)}_{(j+1,\ell-1)}} \ta^\dagger_{j,\ell} \ta_{j+1,\ell-1} \nonumber \\
&\sim e^{i\varphi^a_j -i\varphi^a_{j+1} +i\pi x/a_0 +i(\pi-\alpha)} \nonumber \\
&\ \ \times \left[ c_1^2 +\frac{c_2^2}{4} \left( e^{2i\pi \brho (2x-a_0) +2i(\theta^a_j+\theta^a_{j+1})} +\textrm{h.c.} \right) \right], \\
&e^{i\calA^{(j,\ell)}_{(j+1,\ell)}} \ta^\dagger_{j,\ell} \ta_{j+1,\ell} \nonumber \\
&\sim e^{i\varphi^a_j -i\varphi^a_{j+1} +i\pi x/a_0} \nonumber \\
&\ \ \times \left[ c_1^2 +\frac{c_2^2}{4} \left( e^{4i\pi \brho x +2i(\theta^a_j +\theta^a_{j+1})} \right) +\textrm{h.c.} \right]. 
\end{align}
\end{subequations}
Here and hereafter we only keep terms containing the vertex operators $e^{\pm in (\theta^s_j+\theta^s_{j+1})}$, which attribute to the BIQH and Halperin (221) phases. 
Then the string parts are bosonized as 
\begin{subequations}
\begin{align}
\tK^b_{j,\ell-1} \tK^b_{j+1,\ell-1} &\sim \frac{1}{4} e^{2i\pi \brho (x-a_0) +i(\theta^b_j+\theta^b_{j+1}) +i\pi N^b_j} +\textrm{h.c.}, \\
\tK^b_{j,\ell+1} \tK^b_{j+1,\ell} &\sim \frac{1}{4} e^{i\pi \brho (2x+a_0) +i(\theta^b_j +\theta^b_{j+1}) +i\pi N^b_j} +\textrm{h.c.} 
\end{align}
\end{subequations}
Combining these expressions yields the first two terms in Eq.~\eqref{eq:SPTHam1} for $\brho a_0=1/2$ and those in Eq.~\eqref{eq:221Ham} for $\brho a_0=1/6$. 
If we did not incorporate the density operators into the string, the interaction Hamiltonian was equivalently written as 
\begin{align}
H_1 =& \ t' \sum_{j=1}^N \sum_{\ell=1}^L \Bigl[ e^{i\calA^{(j,\ell)}_{(j+1,\ell-1)}} \ta^\dagger_{j,\ell} \ta_{j+1,\ell-1} \tK^b_{j,\ell} \nonumber \\
& \hspace{10pt} \times \tK^b_{j+1,\ell-1} (2\tn^b_{j,\ell-1}-1) \nonumber \\
&+e^{i\calA^{(j,\ell)}_{(j+1,\ell)}} \ta^\dagger_{j,\ell} \ta_{j+1,\ell} \tK^b_{j,\ell} \tK^b_{j+1,\ell} (2\tn^b_{j,\ell}-1) \nonumber \\
& +e^{i\calB^{(j,\ell)}_{(j+1,\ell-1)}} \tb^\dagger_{j,\ell} \tb_{j+1,\ell-1} \tK^a_{j,\ell} \tK^a_{j+1,\ell-1} (2\tn^a_{j+1,\ell}-1) \nonumber \\
& +e^{i\calB^{(j,\ell)}_{(j+1,\ell)}} \tb^\dagger_{j,\ell} \tb_{j+1,\ell} \tK^a_{j,\ell} \tK^a_{j+1,\ell} (2\tn^a_{j+1,\ell}-1) +\textrm{h.c.} \Bigr]. 
\end{align}
Let us consider only the string parts of the first two terms. 
Those are bosonized as  
\begin{subequations}
\begin{align}
& \tK^b_{j,\ell} \tK^b_{j+1,\ell-1} (2\tn^b_{j,\ell-1}-1) \nonumber \\
&\sim \frac{1}{4\pi} \left( e^{i(\theta^b_j+\theta^b_{j+1})+i\pi \brho (2x-3a_0) +i\pi N_j^b} +\textrm{h.c.} \right) \nonumber \\
& +\frac{2\brho a_0-1}{4} \left( e^{i(\theta^b_j+\theta^b_{j+1})+i\pi \brho (2x-a_0) +i\pi N_j^b} +\textrm{h.c.} \right), \\
& \tK^b_{j,\ell} \tK^b_{j+1,\ell} (2\tn^b_{j,\ell}-1) \nonumber \\
&\sim \frac{2\brho a_0 -1+1/\pi}{4} \left( e^{i(\theta^b_j +\theta^b_{j+1}) +2i\pi \brho x +i\pi N^b_j} +\textrm{h.c.} \right). 
\end{align}
\end{subequations}
Combined with Eq.~\eqref{eq:HopBoson}, for $\brho a_0 = 1/2$, this gives the coupling constants just scaled by $1/\pi$ from those of Hamiltonian \eqref{eq:SPTHam2}, and the phase shifts of the cosine potentials are also modified. 
For $\brho a_0 = 1/6$, we obtain the Hamiltonian of the form \eqref{eq:221Ham}, but the coupling constant $f(\alpha)$ is replaced by 
\begin{align}
\tf(\alpha) =& \ \frac{2}{3\pi} \Bigl[ 2(4\pi^2-9\pi+9) -3\sqrt{3}(2\pi-3) \cos \alpha \nonumber \\
&+(8\pi^2-18\pi+9) \sin \alpha \Bigr]^{1/2}.  
\end{align}
Correspondingly, the phase shifts also become more complicated functions in $\alpha$ and $\beta$. 
However, the relative strength between $\tf(\alpha)$ and $\tf(-\alpha)$ still holds the same property as $f(\alpha)$, that is $|f(\alpha)/f(-\alpha)|>1$ for $0 < \alpha < \pi$. 
Therefore, regardless of the different continuum expressions of the Hamiltonian, we obtain the same low-energy physics in both cases. 

\section{Renormalization group analysis} \label{app:RG}

To obtain the phase diagram for general values of the staggered fluxes $\alpha$ and $\beta$, we here analyze the renormalization group (RG) equations of the coupling constants in Eq.~\eqref{eq:SPTHam2}. 
We consider the Hamiltonian, 
\begin{widetext}
\begin{align}
H =& \ \frac{v}{2\pi} \sum_{j=1}^N \sum_{s=a,b} \int dx \left[ \frac{1}{K_s} (\partial_x \theta^s_j)^2 +K_s (\partial_x \varphi^s_j)^2 \right] \nonumber \\
&+\frac{2}{(2\pi a_0)^2} \sum_{j=1}^N \int dx \Biggl[ 
\tg_{1,j} \cos (\phi^1_j-\tphi^1_{j+1}) +\tg_{2,j} \cos (\tphi^1_j-\phi^1_{j+1}) +\tg_{3,j} \cos (\phi^2_j-\tphi^2_{j+1}) +\tg_{4,j} \cos (\tphi^2_j-\phi^2_{j+1}) \nonumber \\
&+\tq_{1,j} \left\{ \sum_{\epsilon= \pm 1} \cos \left( \phi^1_{j-1}-\tphi^1_j +\epsilon (\phi^2_j-\tphi^2_{j+1}) \right) +\sum_{\epsilon= \pm 1} \cos \left( \phi^2_{j-1} -\tphi^2_j +\epsilon (\phi^1_j-\tphi^1_{j+1}) \right) \right\} \nonumber \\
&+\tq_{2,j} \left\{ \sum_{\epsilon= \pm 1} \cos \left( \tphi^1_{j-1} -\phi^1_j +\epsilon (\tphi^2_j-\phi^2_{j+1}) \right) +\sum_{\epsilon= \pm 1} \cos \left( \tphi^2_{j-1} -\phi^2_j +\epsilon(\tphi^1_j-\phi^1_{j+1}) \right) \right\} 
\Biggr].
\end{align}
Here we have neglected the phase shifts in the cosine potentials since it does not affect the present analysis. 
We have also explicitly introduced the stiffness (or the Luttinger parameter) $K_s$ for each species of the original bosonic fields. 
A deviation from $K_s=1$ corresponds to the generation of marginal terms $\partial_x \phi^1 \partial_x \phi^2$ and $\partial_x \tphi^1 \partial_x \tphi^2$ in Eq.~\eqref{eq:ChainBoson2}.
The vertex operators proportional to $\tq_{n,j}$ ($n=1,2$) are not originally included in  Hamiltonian \eqref{eq:SPTHam2}, but those operators are generated by second-order perturbations $\tg_{1,j \pm 1} \tg_{3,j}$ and $\tg_{2,j \pm 1} \tg_{4,j}$ under the RG transformation. 
While similar vertex operators can also be generated by perturbations $\tg_{1,j \pm 1} \tg_{4,j}$ and $\tg_{2,j \pm 1} \tg_{3,j}$, those have nonzero conformal spins $\pm 1$ and do not contribute to the RG equations of the coupling constants $\tg_{n,j}$ ($n=1,\cdots,4$) at the one-loop level. 
Using the standard RG method combined with the operator product expansion \cite{Cardy} and setting $\tg_{n,j} \equiv \tg_n$ and $\tq_{n,j} \equiv \tq_n$, we obtain the RG equations, 
\begin{align}
\begin{split}
\frac{dK_a}{dl} &= \frac{1}{8} (G_3^2+G_4^2) -\frac{K_a^2}{8} (G_1^2+G_2^2) +\frac{1-K_a^2}{2}(Q_1^2+Q_2^2), \\
\frac{dK_b}{dl} &= \frac{1}{8} (G_1^2+G_2^2) -\frac{K_b^2}{8} (G_3^2+G_4^2) +\frac{1-K_b^2}{2}(Q_1^2+Q_2^2), \\
\frac{dG_1}{dl} &= \left( 2-\frac{K_a+K_b^{-1}}{2} \right) G_1 -2Q_1 G_3, \\
\frac{dG_2}{dl} &= \left( 2-\frac{K_a+K_b^{-1}}{2} \right) G_2 -2Q_2 G_4, \\
\frac{dG_3}{dl} &= \left( 2-\frac{K_b+K_a^{-1}}{2} \right) G_3 -2Q_1 G_1, \\
\frac{dG_4}{dl} &= \left( 2-\frac{K_b+K_a^{-1}}{2} \right) G_4 -2Q_2 G_2, \\
\frac{dQ_1}{dl} &= \left( 2-\frac{K_a+K_a^{-1}+K_b+K_b^{-1}}{2} \right)Q_1 -\frac{1}{2} G_1 G_3, \\
\frac{dQ_2}{dl} &= \left( 2-\frac{K_a+K_a^{-1}+K_b+K_b^{-1}}{2} \right)Q_2 -\frac{1}{2} G_2 G_4, 
\end{split}
\end{align}
\end{widetext}
where $G_n = \tg_n/\pi v$ and $Q_n = \tq_n/\pi v$. 
Initial couplings are given by $K_a(0) = K_b(0) =1$, $\tg_1(0) = -\pi^2 t' c_1^2 a_0 g(\alpha)$, $\tg_2(0) = -\pi^2 t' c_1^2 a_0 g(-\alpha)$, $\tg_3(0) = -\pi^2 t' c_1^2 a_0 g(\beta)$, $\tg_4(0) = -\pi^2 t' c_1^2 a_0 g(-\beta)$, and $\tq_1(0) = \tq_2(0) =0$ [see Eq.~\eqref{eq:gandgamma} for the definition of $g(\alpha)$]. 
By inclusion of $Q_n$, pairs of the coupling constants $\{ G_1, G_2 \}$ and $\{ G_3, G_4 \}$ cannot independently flow under the RG transformation and are coupled via $Q_n$. 
Thus the symmetry of the equations under $G_1 \leftrightarrow G_2$ is broken by the asymmetry under $G_3 \leftrightarrow G_4$, and vise versa. 
One can also add other vertex operators generated by second-order perturbations, but they maintain this symmetry and have no essential contributions to the RG flows of $\tg_n$. 

By numerically integrating the RG equations, we find that there exists a nontrivial fixed point $\tg_n = \tg^* \to -\infty$ and $K_s =1$ when the initial couplings satisfy $\tg_1(0)=\tg_4(0)$ and $\tg_2(0)=\tg_3(0)$. 
This initial condition corresponds to the transition lines in the phase diagram of Fig.~\ref{fig:PhaseDiagram}. 
Such a fixed point has also been found in the sliding Luttinger liquid perturbed by the vertex operators of dual fields \cite{Sierra03}. 
Our result can be seen as an extension of their result to a two-component sliding Luttinger liquid. 
A deviation from the above initial condition immediately gives the BIQH phases described by the fixed points $\tg_1^*=\tg_3^*$, $\tg_2^*=\tg_4^*$, and $\tg_1^*/\tg_2^* \to \infty$ or $0$. 

We also remark that for arbitrary initial values of $\tg_n$, we can also find other fixed points $(\tg_1^*, \tg_2^*,\tg_3^*,\tg_4^*) = (\pm \infty, \pm \infty, 0,0)$ or $(0,0, \pm \infty, \pm \infty)$. 
These fixed points describe ordered phases where the bosons condense in one sublattice while the vortices condense in the other sublattice.  
However, these phases are not realized in the parameter space of  Hamiltonian \eqref{eq:SPTHam2}. 

\bibliography{U1SPT_Refs}

\end{document}